\documentclass[preprint,showpacs,showkeys,aps,pra]{revtex4}
\usepackage{tabularx}
\usepackage{dcolumn}
\usepackage{graphics}
\usepackage{bm}
\usepackage[dvips]{graphicx}
\usepackage{tabularx}
\usepackage{amsmath}
\usepackage{amssymb}
\usepackage{longtable}
\usepackage{verbatim}
\usepackage{float}
\usepackage{fancyhdr}
\usepackage{color}
\usepackage {extarrows}

\begin{document}
\pagestyle{fancy}
\rhead{\thepage}
\chead{}
\lhead{}
\cfoot{}
\rfoot{}
\lfoot{}

\title{Properties of Excited Charmed-Bottom Mesons}
\author{Ishrat Asghar}\email{ishrat.2000@gmail.com}\affiliation{Centre For High Energy Physics, Punjab
University, Lahore(54590), Pakistan. }
\author{Faisal Akram}\email{faisal.chep@pu.edu.pk}\affiliation{Centre For High Energy Physics, Punjab
University, Lahore(54590), Pakistan. }
\author{Bilal Masud}\email{bilalmasud.chep@pu.edu.pk}\affiliation{Centre For High Energy Physics, Punjab
University, Lahore(54590), Pakistan. }
\author{M. Atif Sultan}\email{atifsultan.chep@pu.edu.pk}\affiliation{Centre For High Energy Physics, Punjab
University, Lahore(54590), Pakistan. }

\begin{abstract}
 We calculate the spectrum of $B_c$ mesons using a non-relativistic quark potential model. Using the calculated wave functions, we compute the radiative widths of $B_c$ excited states. The strong decay widths are calculated in a modified $^3P_0$ model, assuming harmonic oscillator wave functions. The hadronic transition rates of $B_c$ mesons are calculated using the Kuang-Yan approach. These results are used to determine branching ratios of possible decay channels of several $B_c$ excited states. Calculated branching ratios are then combined with production cross section of $B_c$ states at the LHC to suggest strategies to find missing excited states of $B_c$ mesons.
\end{abstract}

\pacs{12.39.-x, 13.20.-v, 13.25.-k, 12.40.-y}
\keywords{Quark potential model, Spectroscopy of $B_c$ mesons, Decays of $B_c$ mesons,}
\maketitle

\section{Introduction}
$B_c^{+}$ meson is the lowest-mass bound state of a charm quark and a bottom antiquark. This pseudoscalar mesonic ground state has no electromagnetic or strong decays as it cannot annihilate into gluons or photons. It was first observed by the CDF Collaboration at Fermilab through a semi-leptonic decay mode $B_c^{\pm}\rightarrow J/\psi~l^{\pm}\nu$ in $p\overline{p}$ collisions. The measured mass and life time of $B_c$ by CDF were $6.40\pm 0.39\pm 0.13$~GeV and $0.46^{+0.18}_{-0.16}\pm 0.03$~ps respectively \cite{abe-CDF-1998}. It has also been observed by LHCb \cite{aaij-LHCb-2012} and D0 \cite{abazov-D0-2008} experiments through its different decay channels. In 2014 an excited state of $B_c$ meson was observed by ATLAS experiment at LHC through the
decay channel $B_c^{\pm}\rightarrow J/\psi~\pi^{\pm}$ in $pp$ collisions.  The measured mass of the excited state was found to be $6842\pm 4\pm 5$~MeV \cite{aad-ATLAS-2014}, which is considered as the second S-wave state of $B_c$. However, this excited state has not been confirmed by other experiments yet. Excited $B_c$ states below $BD$ threshold ($\approx7144$ MeV) can only decay through radiative and hadronic transitions to $B_c$ ground state, which decay through weak interaction. There are at least two $S$-wave, two $P$-wave, and one $D$-wave $B_c$ multiplets lying below the threshold. Each of these states cascades into $B_c$ ground state through emission of photons and/or pions only. This results into unique experimental signatures through which we can identify them. This is particularly important when a large sample of $B_c$ states is expected to be produced at LHC. To predict event rates of various decay chains of excited $B_c$ states lying below the $BD$ threshold at LHC, we require a knowledge of branching ratios of their electromagnetic and hadronic transitions along with their production cross sections.

There have been many calculations of $B_c$ spectrum using non-relativistic and relativistic quark models ~\cite{ferretti-2018,ebert-2003,yuan-2012,godfrey-2004-Bc,parkash-2017,bhat-2017}. The electromagnetic transitions of $B_c$ are predicted in Refs.~\cite{godfrey-2004-Bc,ebert-2003,parkash-2017,bhat-2017,hady-2005} and hadronic transitions are calculated in \cite{godfrey-2004-Bc,hong-2010}. In Ref. \cite{ferretti-2018}, the open-flavor strong decay widths of $B_c$ mesons are predicted in the $^3P_0$ model. Ref. \cite{ferretti-2018} studied only strong decays to pairs of S-wave mesons for many open flavor states, using the same value of the harmonic oscillator parameter $\beta$ for all the flavor states. The present work  provides a comprehensive theoretical study of $B_c$ mesons properties: Here we calculate the spectrum, radiative transitions (E1 and M1), hadronic transitions and strong decays of $B_c$ mesons following a consistent approach. A non-relativistic potential quark model is used to explain the mass spectrum of $B_c$ mesons. The wavefunctions computed through this model are then used to find the decay widths of their E1 and M1 transitions.
The hadronic transitions of $B_c$ mesons are estimated by using the Kuang-Yan approach. We use the $^3P_0$ model to study the open-flavor strong decays of $B_c$ states. In comparison to Ref. \cite{ferretti-2018}, we have calculated strong decay widths of all angularly excited $B_c$
states while using different values of $\beta$ for different flavor states. We combine radiative, hadronic and strong widths to predict the branching ratios of all possible decay channels of several $B_c$ excited states. The branching ratios of radiative and hadronic transitions of $B_c$ excited states lying below $BD$ threshold are combined with their predicted production cross sections at LHC energy to provide estimates of event rates of their possible decay chains through which these states can be identified in the experimental data.

The organization of the paper is as follows. First, we describe the potential model used to calculate the mass spectrum of charm-bottom mesons. In Sec. \ref{sect:open-flavor-strong-decays}, we review the $^3P_0$ decay model and evaluate the strong decay amplitudes. E1 and M1 radiative transitions are calculated in Sec. \ref{sect:radiative-transitions}. This is followed by the estimates in Sec. \ref{sect:hadronic transitions} of hadronic transitions based on the Kuang-Yan approach. We discuss the best strategies for searching the excited $B_c$ states lying below the $BD$ threshold in Sec. \ref{sec:experimental signatures}, while our concluding remarks are given in Sec. \ref{conclusions}.
\section{Mass spectrum} \label{sect:mass-spectrum}
In this section we give the mass predictions of the non-relativistic quark model for charm-bottom mesons. We use the standard ``Coulomb+linear" potential, and spin-dependent corrections generated from vector gluon exchange and an effective scalar confinement interaction. The potential used in this paper is given by~\cite{barnes-2005}
\begin{eqnarray}\label{potential}
V_{q\overline{q}}(r) &=& -\frac{4\alpha_s}{3r} + br + \frac{32\pi \alpha_s}{9m_qm_{\overline{q}}}(\frac{\sigma}{\sqrt{\pi}})^3 e^{-\sigma^2r^2}\textbf{S}_q\cdot \textbf{S}_{\overline{q}}+ \frac{4\alpha_s}{m_q m_{\overline{q}}r^3} T \nonumber \\
 && + \left( \frac{\textbf{S}_q}{4 m_q^2} + \frac{\textbf{S}_{\overline{q}}}{4 m_{\overline{q}}^2}\right) \cdot \textbf{L} \left(\frac{4 \alpha_s}{3 r^3} - \frac{b}{r}\right) + \frac{\textbf{S}_q + \textbf{S}_{\overline{q}}}{2 m_q m_{\overline{q}}} \cdot \textbf{L} \, \frac{4 \alpha_s}{3 r^3}.
\end{eqnarray}
Here $\alpha_s$ is the strong coupling constant, $b$ is the string tension,
and $T$ is the tensor operator
\begin{equation}
T = \textbf{S}_q\cdot \hat r \, \textbf{S}_{\overline{q}}\cdot \hat r - \frac{1}{3} \textbf{S}_q\cdot \textbf{S}_{\overline{q}},
\end{equation}
with diagonal matrix elements given by
\begin{equation}
 T= \left\{
      \begin{array}{ll}
        -\frac{L}{6(2L+3)}&\hspace{0.4cm} J=L+1 \\
       \frac{1}{6}&\hspace{0.4cm} J=L \\
        -\frac{(L+1)}{6(2L-1)}&\hspace{0.4cm} J=L-1.
      \end{array}
    \right.
\end{equation}
The strong coupling constant $\alpha_s$ used in this potential for $B_c$ mesons is taken to be $0.4$. This value was obtained by our fit to the masses of two experimentally known states of $B_c$ mesons (these are listed in $4^{th}$ column of Table \ref{Bc table-1}).
The parameters $\sigma=0.84$~GeV, $b=0.0945$ GeV$^2$, $m_{u/d}=0.325$~GeV and $m_s=0.422$~GeV were obtained from fits to light mesons~\cite{ishrat-2018}, and $m_c=1.4794$~GeV is from charmonium sector. Finally $m_b=4.825$~GeV is taken from Ref. \cite{nosheen-2017}.

The meson states with quark and antiquark of unequal mass are not charge conjugation eigenstates. Therefore, states with $J=L$ and $S=0,1$, i.e., $|n\;^1L_J\rangle$ and $|n\;^3L_J\rangle$ can mix via spin-orbit interaction. For example $^1P_1$ and $^3P_1$ states can mix through the following linear combination
\begin{eqnarray}\label{mixed-states}
  |B(1\;P_1)\rangle=+\cos(\phi_{1P})|1\;^1P_1\rangle+\sin(\phi_{1P})|1\;^3P_1\rangle \nonumber\\
  |B(1\;P^{'}_1)\rangle=-\sin(\phi_{1P})|1\;^1P_1\rangle+\cos(\phi_{1P})|1\;^3P_1\rangle,
\end{eqnarray}
\noindent where $\phi_{1P}$ is the mixing angle. In the heavy quark limit $m_Q\rightarrow \infty$ the mixing angles become~\cite{cahn-2003}
\begin{equation}\label{mixing angle}
  \phi_{m_Q\rightarrow \infty}=\arctan(\sqrt{\frac{L}{L+1}}).
\end{equation}
This implies $\phi_{nP}=35.3^{\circ}$, $\phi_{nD}=39.2^{\circ}$ and $\phi_{nF}=40.89^{\circ}$. The mixing angles in heavy quark limit for $1D$ and $1F$ states are close to those produced by Ref. \cite{godfrey-2004-Bc} whereas for $1P$ state it is slightly different. The spectrum of $B_c$ states obtained by solving the radial Schr\"{o}dinger equation through the shooting method~\cite{atif-2014} is reported in Tables \ref{Bc table-1} and \ref{Bc table-2}.
\begin{table}[H]
\centering
\renewcommand{\arraystretch}{0.6}
\caption{Masses of ground and excited states of $B_c$ mesons. The mixed states $P_1-P'_1$, $D_2-D'_2$ and $F_3-F'_3$ and their mixing angles $\phi_{nP}$, $\phi_{nD}$ and $\phi_{nF}$ are defined according to Eq.~(\ref{mixing angle}). The SHO $\beta$ values are listed in the last column which are obtained by fitting SHO wave functions to the quark model wavefunctions.}
\begin{tabular}{c c c c c c c}
\hline\hline
\hspace{0.3cm} nL \hspace{0.3cm} & \hspace{0.3cm} Meson \hspace{0.3cm} & \hspace{0.2cm} Our calculated mass \hspace{0.2cm} & \hspace{0.3cm} Expt. mass \cite{PDG-2016}& \hspace{0.3cm} $\beta$ \hspace{0.3cm}\\
\hspace{0.3cm}  \hspace{0.3cm} & \hspace{0.3cm}  \hspace{0.3cm} & \hspace{0.2cm} (GeV) \hspace{0.2cm} & \hspace{0.3cm} (MeV)& \hspace{0.3cm} (GeV) \hspace{0.3cm}\\
   \hline
1S & $B_c(1^1S_0)$  & 6.318 & $6274.9\pm 0.8$ & 0.653\\
   & $B_c(1^3S_1)$  & 6.336 & ...          & 0.634\\
   \hline
2S & $B_c(2^1S_0)$  & 6.741 & $6842\pm 4$  & 0.515\\
   & $B_c(2^3S_1)$  & 6.747 & ...          & 0.508\\
\hline
3S & $B_c(3^1S_0)$  & 7.014 & ...          & 0.442\\
   & $B_c(3^3S_1)$  & 7.018 & ...          & 0.439\\
\hline
4S & $B_c(4^1S_0)$  & 7.239 &...           &0.402\\
   & $B_c(4^3S_1)$  & 7.242 & ...          &0.401\\
\hline
1P & $B_c(1^3P_0)$  & 6.631 & ...          &0.468\\
   & $B_c(1^3P_2)$  & 6.665 & ...          & 0.468\\
   & $B_c(1\;P_1)$  & 6.650 & ...          & 0.471,0.468\\
   & $B_c(1\;P'_1)$ & 6.656 & ...          & 0.471,0.468\\
   & $\phi_{1P}$    &$35.3^\circ$       &  \\
\hline
2P & $B_c(2^3P_0)$  & 6.915 & ...          & 0.428\\
   & $B_c(2^3P_2)$  & 6.946 &...           & 0.428 \\
   & $B_c(2\;P_1)$  & 6.930 & ...          & 0.430,0.428\\
   & $B_c(2\;P'_1)$ & 6.939 & ...          & 0.430,0.428\\
   & $\phi_{2P}$    &$35.3^\circ$       & ...          \\
\hline
3P & $B_c(3^3P_0)$  & 7.147 & ...          & 0.395\\
   & $B_c(3^3P_2)$  & 7.176 &...           & 0.395 \\
   & $B_c(3\;P_1)$  & 7.162 & ...          & 0.397,0.395 \\
   & $B_c(3\;P'_1)$ & 7.168 & ...          & 0.397,0.395\\
   & $\phi_{3P}$    &$35.3^\circ$       & ...   \\
\hline
4P & $B_c(4^3P_0)$  & 7.350 & ...          &0.373        \\
   & $B_c(4^3P_2)$  & 7.379 &...           &0.373         \\
   & $B_c(4\;P_1)$  & 7.364 & ...          & 0.374,0.373 \\
   & $B_c(4\;P'_1)$ & 7.373 & ...          & 0.374,0.373 \\
   & $\phi_{4P}$    &$35.3^\circ$       & ...          &\\
\hline
1D & $B_c(1^3D_1)$  & 6.841 & ...        &0.417        \\
   & $B_c(1^3D_3)$  & 6.847 &...         &0.417         \\
   & $B_c(1\;D_2)$  & 6.845 &...         &0.417,0.417 \\
   & $B_c(1\;D'_2)$ & 6.845 & ...        &0.417,0.417 \\
   & $\phi_{1D}$    &$39.2^\circ$       & ...        \\
\hline
2D & $B_c(2^3D_1)$  & 7.080 & ...               & 0.395          \\
   & $B_c(2^3D_3)$  & 7.087 &...             & 0.395         \\
   & $B_c(2\;D_2)$  & 7.084 &...                & 0.395,0.395  \\
   & $B_c(2\;D'_2)$ & 7.084 & ...               & 0.395,0.395  \\
   & $\phi_{2D}$    &$39.2^\circ$       & ...          \\
\hline
3D & $B_c(3^3D_1)$  & 7.289 & ...           &0.374       \\
   & $B_c(3^3D_3)$  & 7.296 &...            &0.374      \\
   & $B_c(3\;D_2)$  & 7.293 &...            &0.374,0.374         \\
   & $B_c(3\;D'_2)$ & 7.293 & ...           &0.374,0.374          \\
   & $\phi_{3D}$    &$39.2^\circ$       & ...           \\
\hline
4D & $B_c(4^3D_1)$  & 7.478 & ...          &0.357        \\
   & $B_c(4^3D_3)$  & 7.489 &...           &0.357        \\
   & $B_c(4\;D_2)$  & 7.482 &...           &0.358,0.357        \\
   & $B_c(4\;D'_2)$ & 7.483 & ...          &0.358,0.357       \\
   & $\phi_{4D}$    &$39.2^\circ$       & ...           \\
  \hline\hline
\end{tabular}
\label{Bc table-1}
\end{table}

\begin{table}[H]
\centering
\renewcommand{\arraystretch}{0.6}
\caption{Masses of ground and excited states of $B_c$ mesons (continued).}
\begin{tabular}{c c c c c c c}
\hline\hline
\hspace{0.3cm} nL \hspace{0.3cm} & \hspace{0.3cm} Meson \hspace{0.3cm} & \hspace{0.2cm} Our calculated mass \hspace{0.2cm} & \hspace{0.3cm} Expt. mass \cite{PDG-2016}& \hspace{0.3cm} $\beta$ \hspace{0.3cm}\\
\hspace{0.3cm}  \hspace{0.3cm} & \hspace{0.3cm}  \hspace{0.3cm} & \hspace{0.2cm} (GeV) \hspace{0.2cm} & \hspace{0.3cm} (MeV)& \hspace{0.3cm} (GeV) \hspace{0.3cm}\\
   \hline
1F & $B_c(1^3F_2)$  & 6.9972  & ...    & 0.390           \\
   & $B_c(1^3F_4)$  & 6.9967  &...     & 0.390              \\
   & $B_c(1\;F_3)$  & 6.994   &...     & 0.390,0.390              \\
   & $B_c(1\;F'_3)$ & 7.001   &...     & 0.390,0.390        \\
   & $\phi_{1F}$    &$40.89^\circ$       & ...      \\
\hline
2F & $B_c(2^3F_2)$  & 7.2121 & ...     &0.375             \\
   & $B_c(2^3F_4)$  & 7.2126 &...      &0.375              \\
   & $B_c(2\;F_3)$  & 7.211   &...     &0.375,0.375               \\
   & $B_c(2\;F'_3)$ & 7.214   &...     &0.375,0.375             \\
   & $\phi_{2F}$    &$40.89^\circ$       & ...           \\
  \hline\hline
\end{tabular}
\label{Bc table-2}
\end{table}
\section{Open Flavor Strong Decays}
\label{sect:open-flavor-strong-decays}
In the $^3P_0$ model, the open-flavor strong decay of a meson takes place through production of a light $q\bar{q}$ pair ($q=u,d,s$) with vacuum quantum numbers ($J^{PC}=0^{++}$). The interaction Hamiltonian for the $^3P_0$ model is obtained from the nonrelativistic limit of
\begin{equation}\label{hamiltonian}
H_I=2 m_q \gamma\int d^3\textbf{x}\;\overline{\psi}_q(\textbf{x}) \psi_q(\textbf{x}),
\end{equation}
where $\gamma$ is a dimensionless pair production strength. The pair-production strength parameter $\gamma$ is fitted to strong decay data. In the original $^3P_0$ model introduced by Micu \cite{micu-1969}, new $q\overline{q}$ pair is produced by a constant pair production amplitude $\gamma$. Some variants of the $^3P_0$ model include an effective pair production strength $(\gamma^{eff})$ that
suppresses heavy $q\overline{q}$ pair production \cite{kalasnikova-2005,ferretti-2012}.

The $^3P_0$ model has been successfully applied to strong decays of light mesons \cite{barnes-1997}, strange mesons \cite{blundell-1996,barnes-2003}, charmonium states \cite{barnes-2005}, bottomonium states \cite{ferretti-2014,godfrey-2015}, open-charm \cite{close-2005,godfrey-2016-open-charm} and open-bottom sectors \cite{yuan-2014-open-bottom,godfrey-2016-open-bottom,ishrat-2018,ferretti-2018}. In this study, we have computed strong decay widths of kinematically allowed open-flavor decay modes of all the $B_c$ states mentioned in Table \ref{Bc table-1} and \ref{Bc table-2} using the $^3P_0$ model. The interaction Hamiltonian for the $^3P_0$ model can be written in terms of the creation operators as
\begin{equation}\label{interaction-hamiltonian}
  H_I=2m_q\gamma \int d^3k[\overline{u}(\mathbf{k},s)v(\mathbf{-k},\overline{s})]b^{\dag}(\textbf{k},s)d^{\dag}(-\textbf{k},\overline{s}),
\end{equation}
where $b^{\dag}$ and $d^{\dag}$ are the creation operators for quark and antiquark respectively. The pair production strength factor $\gamma = 0.35$ is obtained from a fit of known strong decay widths of the $c\overline{c}$ states above open-charm threshold~\cite{barnes-2005}. In this work, we use a modified version of pair production strength that replaces $\gamma$ with
\begin{equation}
\label{eq:eff}
  \gamma^{\textmd{eff}}=\frac{m_{u/d}}{m}\;\gamma,
\end{equation}
where $m$ is the mass of the produced quark \cite{kalasnikova-2005,ferretti-2012}. This mechanism suppresses those diagrams in which a heavy $q\overline{q}$ pair is created.

\begin{figure}
  \centering
  \includegraphics[width=10cm]{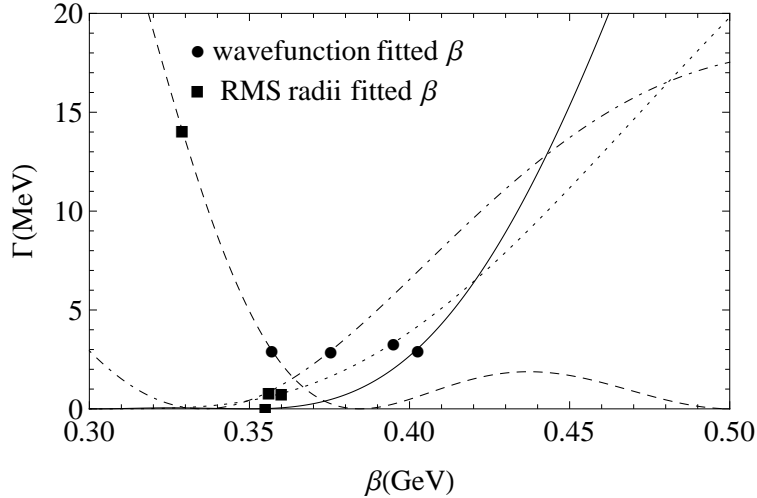}
  \caption{Plots of partial widths of some of the strong decays of $B_c$ mesons in the $^3P_0$ model as a function of $\beta$ for the initial meson. Solid: $B_c(4\;^1S_0)\rightarrow DB^*$, dotted: $B_c(3\;^3P_0)\rightarrow DB$, dashed: $B_c(4\;^3D_1)\rightarrow DB$, dot-dashed: $B_c(2\;^3F_2)\rightarrow DB$. Disk and rectangular marks on each curve corresponds to the $\beta$ values obtained by fit to numerically calculated wave functions and to rms radii respectively.}
  \label{beta-dependence}
\end{figure}
With the $^3P_0$ model, we use simple harmonic oscillator (SHO) wavefunctions and SHO scale $\beta$ is taken as parameter of the model.
In refs.~\cite{close-2005,godfrey-2016-open-charm,yuan-2014-open-bottom,godfrey-2016-open-bottom}, the $\beta$ parameter was obtained by equating the root mean squared (RMS) radius of a harmonic oscillator wavefunction to the RMS radius of the quark model wavefunction. In this work, we fit $\beta$ of SHO wavefunctions to the wavefunctions obtained by numercially solving Schr\"{o}dinger equation for the potential given in Eq. (\ref{potential}). The resulting $\beta$ values, that are more accurate, are listed in  Tables \ref{Bc table-1}-\ref{parameters for open-charm and open-botom mesons} for the initial $B_c$ mesons, and the final $D$, $D_s$, $B$ and $B_s$ mesons appearing in strong decays of $B_c$ excited states. These two methods of finding $\beta$ are compared in our earlier work \cite{ishrat-2018}. In Fig. \ref{beta-dependence} we show the dependence of strong decay widths of few decay channels on the value of SHO parameter $\beta$. Disk and rectangular marks on each curve corresponds to the $\beta$ values obtained by fit to numerically calculated wave functions and to rms radii respectively. These plots show that fitted $\beta$ values lie in the sensitive regions of the curves, which implies that, an accurate method of determining the values of $\beta$ parameter can significantly improve the results.

\begin{figure}
  \centering
  \includegraphics[width=10cm]{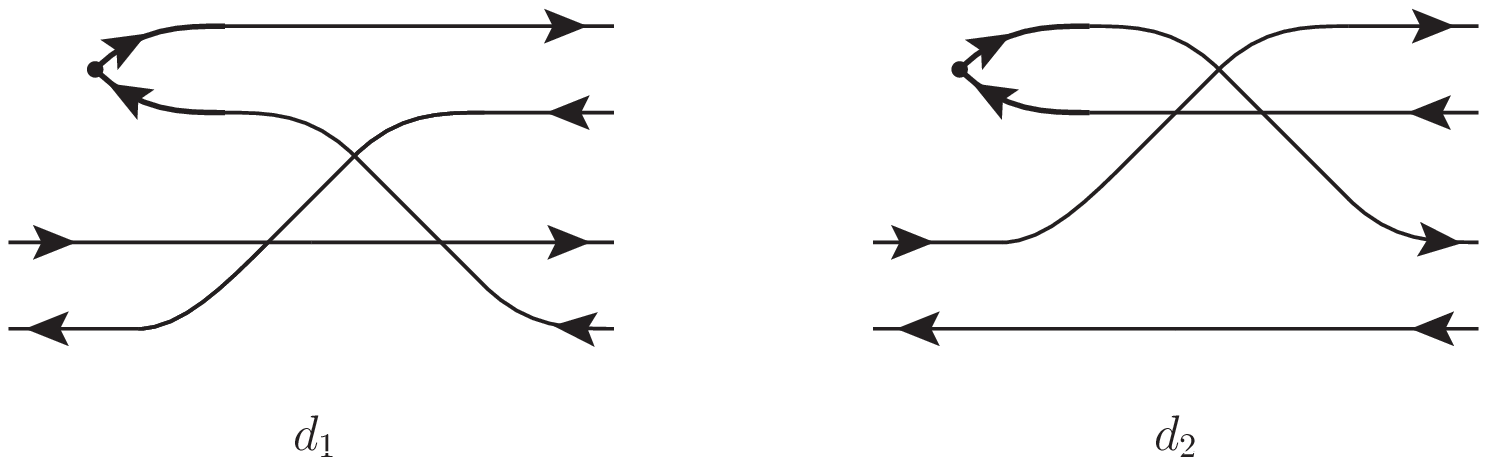}
  \caption{Decay diagrams in the $^3P_0$ model.}
  \label{decaypics}
\end{figure}

To calculate the decay rate of a process $A\rightarrow B+C$, we evaluate the matrix element $\langle BC|H_I|A\rangle$ by using interaction Hamiltonian of Eq. (\ref{interaction-hamiltonian}). In general two different diagrams, shown in Fig. \ref{decaypics}, contribute to the matrix element $\langle BC|H_I|A\rangle$.
Using the relativistic phase space factor from Ref.~\cite{ackleh-1996} and performing the angular integration gives
\begin{equation}\label{gammaTOT}
\Gamma_{A\rightarrow BC}=2\pi\frac{P E_BE_C}{M_A}\sum_{LS} |\mathcal{M}_{LS}|^2,
\end{equation}
where $P=|\textbf{P}_B|=|\textbf{P}_C|$, $M_A$ is the mass of the initial meson, and $E_B=\sqrt{M_B^2+P^2}$ and $E_C=\sqrt{M_C^2+P^2}$ are the energies of the final mesons $B$ and $C$ respectively. Where available, we use experimental masses~\cite{PDG-2016} of $B_c$ mesons; otherwise we use the theoretical masses given in Tables \ref{Bc table-1} and \ref{Bc table-2}. The masses of the final state mesons $D$, $D_s$, $B$ and $B_s$ are reported in Table \ref{parameters for open-charm and open-botom mesons}.
The detailed formulism to calculate the matrix element $\langle BC|H_I|A\rangle$ and decay amplitude $\mathcal{M}_{LS}$ is described in our earlier work \cite{ishrat-2018}.

\begin{table}[h!]
\centering
\renewcommand{\arraystretch}{0.7}
\caption{Masses and SHO parameter ($\beta$) for open-charm and open-bottom mesons used in our strong decay width calculations. The experimental masses are taken from PDG \cite{PDG-2016}. The SHO $\beta$ values are listed in the second column which we obtain by the fit of SHO wave function to the numerical solutions of Schr\"{o}dinger equation.}
\begin{tabular}{c c c c c c c }
\hline
\hline
Meson\hspace{.1 in}& \hspace{.3 in} $\beta$(GeV)\hspace{.3 in} &Expt. Mass \cite{PDG-2016}(MeV) & Mass used in calculations(GeV) \\
\hline\hline
$D$         & 0.442 & $D^{\pm}=1869.58\pm0.09$  & 1.867 \\
            &       & $D^0=1864.83\pm0.08$      \\
$D^*$       & 0.338 & $D^{*\pm}=2010.26\pm0.05$  & 2.007 \\
            &       & $D^{*0}=2006.85\pm0.05$      \\
\hline
$D_s$       & 0.463 & $D_s^{\pm}=1968.27\pm0.10$ & 1.968   \\
$D_s^*$     & 0.369 & $D_s^{*\pm}=2112.1\pm0.4$   & 2.112 \\
\hline
$B$         & 0.405 & $B^{\pm}=5279.31\pm0.15$  & 5.279 \\
            &       & $B^0=5279.62\pm0.15$      \\
$B^*$       & 0.372 & $B^*=5324.65\pm0.25$ & 5.325 \\
\hline
$B_s$       & 0.429 & $B_s^0=5366.82\pm0.22$ & 5.367 \\
$B_s^*$     & 0.401 & $B_s^{*0}=5415.4_{-1.5}^{+1.8}$& 5.415 \\
\hline
\end{tabular}
\label{parameters for open-charm and open-botom mesons}
\end{table}

\section{Radiative Transitions of $B_c$ Mesons}
\label{sect:radiative-transitions}

\subsection{E1 Radiative Transitions}
E1 radiative partial widths are computed  with the following expression~\cite{kwong-1988}
\begin{equation}
  \Gamma(n^{2S+1}L_J\rightarrow n'^{2S'+1}L'_{J'}+\gamma)=\frac{4}{3}\langle e_Q\rangle^2\alpha \omega^3 C_{fi} \delta_{SS'}|\langle n'^{2S'+1}L'_{J'}|r| n^{2S+1}L_J\rangle|^2 \frac{E_f}{M_i},
  \label{e1}
\end{equation}
where
\begin{equation}
  \langle e_Q\rangle=\frac{m_{b}\;e_c-m_c\;e_{\overline{b}}}{m_b+m_c}.
\end{equation}
Here $e_c$ and $e_{\overline{b}}$ are the electric charges of the charm quark and bottom anti-quark in units of $|e|$, $m_b$ and $m_c$ are the constituent masses of the charm and bottom quarks, $\alpha$ is the fine structure constant, $\omega$ is the final photon energy, $M_i$ is mass of the initial meson, and $E_f$ is the energy of the final state. Finally, the angular matrix element $C_{fi}$ is given by
\begin{equation}
 C_{fi}={\text{max}}(L,L')\, (2J'+1)
                            \begin{Bmatrix}
                             L' & J' & S \\
                             J & L & 1 \\
                           \end{Bmatrix}^2. \nonumber
\end{equation}
Eq. \ref{e1} is the result of Ref.~\cite{kwong-1988} except for our inclusion of the relativistic phase space factor $E_f/M_i$ from Refs. \cite{barnes-2005,close-2005}. The matrix elements $\langle n'^{2S'+1}L'_{J'}|r| n^{2S+1}L_J\rangle$ are obtained using the quark model wavefunctions obtained in Sec. \ref{sect:mass-spectrum}.
Wavefunction corrections due to perturbative spin-dependent interactions were neglected in this computation, as in Refs. \cite{barnes-2005,close-2005}.
Results for E1 radiative transitions for $B_c$ mesons are given in Tables \ref{results-1} - \ref{results-8} along with the matrix elements
so that an interested reader can reproduce our results.

\subsection{M1 Radiative Transitions}

The M1 radiative partial widths are evaluated using the following expression~\cite{novikov-1978}
\begin{eqnarray}
\Gamma(n\;^{2S+1}L_J\rightarrow n'\;^{2S'+1}L_{J'}+\gamma) &=& \frac{\alpha}{3}\omega^3\, (2J'+1)\, \delta_{S,S'\pm1} \nonumber \\
  && \hspace{-2cm} \cdot \left|\frac{e_c}{m_c}\langle f|j_0(\frac{m_b}{m_c+m_b}\omega r)|i\rangle
-\frac{e_{\overline{b}}}{m_b}\langle f|j_0(\frac{m_c}{m_c+m_b}\omega r)|i\rangle\right|^2,
\end{eqnarray}
where $j_0(x)$ is a spherical Bessel function. The definitions of the other parameters are the same as in the E1 radiative transitions. The results for M1 radiative transitions for $B_c$ mesons are given in Tables \ref{results-1} - \ref{results-8}.

\section{E1-E1 Hadronic Transitions}
\label{sect:hadronic transitions}

Hadronic transitions are needed to estimate branching ratios and the event rates of decay chains of $B_c$ states lying below $BD$ threshold. The differential rate for E1-E1 hadronic transitions from an initial meson state $\Phi'$ to the final meson state $\Phi$ and a system of light hadrons $h$ is given by~\cite{yan-1980,kuang-1981}
\begin{equation}
\frac{d\Gamma}{d\mathcal{M}^2}[\Phi' \rightarrow \Phi+~h]=(2J+1)\sum_{k=0}^2\begin{Bmatrix}
                             k & l' & l \\
                             s & J & J' \\
                           \end{Bmatrix}^2A_k(l',l),
\label{e1e1}
\end{equation}
where $J'$, $J$ are the total angular momentum and $l'$, $l$ are the orbital angular momentum of initial and final meson states respectively, $\{^{...}_{...}\}$ is a 6-j symbol, $\mathcal{M}^2$ is the invariant mass squared of the light hadron system and $A_k(l',l)$ are the reduced matrix elements~\cite{yan-1980}. Here we use scaling argument to predict the hadronic rates for $c\bar{b}$ mesons using measured rates of $c\bar{c}$ and/or $b\bar{b}$ as input. When measured rates are not available, we use predicted rates of hadronic transition of $b\bar{b}$ states. The scaling law for E1-E1 hadronic transitions is given by~\cite{yan-1980,godfrey-2015}
\begin{equation}
\frac{\Gamma(c\bar{b})}{\Gamma(Q\bar{Q})}=\frac{\langle r^2(c\bar{b})\rangle^2}{\langle r^2(Q\bar{Q})\rangle^2}\;p,
\end{equation}
where $\langle r^2(Q\bar{Q})\rangle$ is the expectation value of
the square of the interquark separation and $p$ is the phase space factor depending on the masses of initial and final states. The phase space factors for $A_0$ and $A_2$ reduced matrix elements are $\frac{G(\Phi'(c\bar{b}) \rightarrow \Phi(c\bar{b})\;\pi\pi)}{G(\Phi'(Q\bar{Q}) \rightarrow \Phi(Q\bar{Q})\;\pi\pi)}$ and $\frac{H(\Phi'(c\bar{b}) \rightarrow \Phi(c\bar{b})\;\pi\pi)}{H(\Phi'(Q\bar{Q}) \rightarrow \Phi(Q\bar{Q})\;\pi\pi)}$ respectively, where $G$ and $H$ are defined in Ref.~\cite{kuang-1981}.

For the $B_c(2S)\rightarrow B_c(1S)+\pi\pi$ reduced rates, we rescale the measured reduced rates of $\psi(2 ^3S_1)\rightarrow \psi(1 ^3S_1)+\pi\pi$ and $\Upsilon(2 ^3S_1)\rightarrow \Upsilon(1 ^3S_1)+\pi\pi$~\cite{PDG-2016} and take their average value. The reduced rates for
$B_c(3S)\rightarrow B_c(1S)+\pi\pi$ are obtained by rescaling the corresponding measured reduced rates of $b\overline{b}$~\cite{PDG-2016}. There are 16 possible $B_c(2P)\rightarrow B_c(1P)+\pi\pi$ hadronic transitions, which can be expressed in terms of three reduced rates $A_0(1,1)$, $A_1(1,1)$ and $A_2(1,1)$ using Eq. \ref{e1e1}. In the soft-pion limit $A_1(l',l)$ contributions are suppressed,  so we take $A_1(1,1)=0$. $A_0(1,1)$ and $A_2(1,1)$ are obtained by rescaling the reduced rates of $\Upsilon(2 ^3P_0)\rightarrow \Upsilon(1 ^3P_0)+\pi\pi$ and $\Upsilon(2 ^3P_2)\rightarrow \Upsilon(1 ^3P_1)+\pi\pi$ predicted by Godfrey and Moats~\cite{godfrey-2015}.
For calculating the hadronic transitions for $B_c(1D)\rightarrow B_c(1S)+\pi\pi$ and $B_c(1F)\rightarrow B_c(1P)+\pi\pi$, we rescale the reduced rates of $\Upsilon(1 ^3D_1)\rightarrow \Upsilon(1 ^3S_1)+\pi\pi$ and $\Upsilon(1 ^3F_2)\rightarrow \Upsilon(1 ^3P_0)+\pi\pi$ which are also taken from Ref.~\cite{godfrey-2015}. The scaling factors and reduced rates between $c\overline{b}$ mesons are given in Table \ref{table:scale-factors}. These reduced rates are used to determine the $c\bar{b}$ hadronic transitions which are summarized in Tables \ref{table:haronic-rates-1} - \ref{table:haronic-rates-2}. To calculate the transition rates for mixed states, mixing angles are incorporated, with the values taken from Table \ref{Bc table-1} and \ref{Bc table-2}.

\begin{table}[H]
\tabcolsep=1pt\fontsize{9.9}{9.9}\selectfont
\centering
\caption{Estimates of reduced rates of hadronic transitions between $c\bar{b}$ mesons.}
\begin{tabular}{c c c c c}
\hline\hline
Transition\hspace{.1 in}&\hspace{.1 in} $(Q\overline{Q})$: rate \hspace{.1 in}&\hspace{.1 in} $\langle r^2(c\overline{b})\rangle/\langle r^2(Q\overline{Q})\rangle$ \hspace{.1 in} & Our reduced $c\overline{b}$ rate \hspace{.1 in}& Reduced $c\overline{b}$ rate~\cite{godfrey-2004-Bc} \hspace{.1 in}\\
&(keV)& &(keV)&(keV) \\
\hline
$2^3S_1\rightarrow 1^3S_1+ \pi\pi$   & $(c\overline{c})$:$155.84\pm5.2~^a$ & 0.94 & $A_0(0,0)=61.59\pm 2.1$ & $A_0(0,0)=82\pm 8$ \\
                                     & $(b\overline{b})$:$8.46\pm 0.7~^a$   & 2.36 & $A_0(0,0)=26.06\pm 2.2$ & $A_0(0,0)=33\pm 5$ \\
                                     & average:                  &      & $43.83\pm 2.2$ & $57\pm7$            \\
$3^3S_1\rightarrow 1^3S_1+ \pi\pi$   & $(b\overline{b})$:$1.34\pm 0.12~^a$   & 2.24 & $A''_0(0,0)=2.08\pm 0.19$ & $A''_0(0,0)=4.2\pm0.6$ \\
$2^3P_0\rightarrow 1^3P_0+ \pi\pi$   & $(b\overline{b})$:0.44~$^b$     & 2.23 & $A_0(1,1)=1.82$ & $A_0(1,1)=2.92$ \\
$2^3P_2\rightarrow 1^3P_1+ \pi\pi$   & $(b\overline{b})$:0.23~$^b$     & 2.22 & $A_2(1,1)=0.328$ & $A_2(1,1)=0.164$ \\
$1^3D_1\rightarrow 1^3S_1+ \pi\pi$   & $(b\overline{b})$:0.14~$^b$     & 2.23 & $A_2(2,0)=1.183$ & $A_2(2,0)=21$ \\
$1^3F_2\rightarrow 1^3P_0+ \pi\pi$   & $(b\overline{b})$:$1.8\times 10^{-3}~^b$     & 2.19 & $A_2(3,1)=0.114$ & ... \\
\hline\hline
\end{tabular}\\
\label{table:scale-factors}
\begin{flushleft}
\hspace{0.1cm}$^a$From PDG Ref.~\cite{PDG-2016}.\\
\hspace{0.1cm}$^b$From Ref.~\cite{godfrey-2015}.
\end{flushleft}
\end{table}

\begin{table}[H]
\centering
\caption{Rates for the E1-E1 hadronic transitions of 2S, 3S and 2P states of $B_c$ mesons.}
\begin{tabular}{c c c}
\hline\hline
Transition\hspace{.1 in}&\hspace{.1 in} Expression for the rate \hspace{.1 in}&\hspace{.1 in} the $c\overline{b}$ rate (keV) \hspace{.1 in}\\
\hline
$2^1S_0\rightarrow 1^1S_0+ \pi\pi$   & $A_0(0,0)$ &  10.68\\
$2^3S_1\rightarrow 1^3S_1+ \pi\pi$   & $A_0(0,0)$ &  3.99\\
$3^1S_0\rightarrow 1^1S_0+ \pi\pi$   & $A''_0(0,0)$ & 0.97 \\
$3^3S_1\rightarrow 1^3S_1+ \pi\pi$   & $A''_0(0,0)$ & 0.81 \\
\hline
$2^3P_0\rightarrow 1\;P_1+ \pi\pi$   & $\frac{1}{3}A_1(1,1)$ & 0 \\
$2^3P_0\rightarrow 1\;P'_1+ \pi\pi$  & $\frac{1}{3}A_1(1,1)$ & 0 \\
$2^3P_0\rightarrow 1^3P_0+ \pi\pi$   & $\frac{1}{3}A_0(1,1)$ & 0.002 \\
\hline
$2^3P_2\rightarrow 1^3P_2+ \pi\pi$   & $\frac{1}{3}A_0(1,1)+\frac{1}{4}A_1(1,1)+\frac{7}{60}A_2(1,1)$ & 0.0007 \\
$2^3P_2\rightarrow 1\;P_1+ \pi\pi$   & $\frac{1}{12}A_1(1,1)+\frac{3}{20}A_2(1,1)$ & 0.0002 \\
$2^3P_2\rightarrow 1\;P'_1+ \pi\pi$  & $\frac{1}{12}A_1(1,1)+\frac{3}{20}A_2(1,1)$ & 0.0002 \\
$2^3P_2\rightarrow 1^3P_0+ \pi\pi$   & $\frac{1}{15}A_2(1,1)$ & 0.0005 \\
\hline
$2\;P_1\rightarrow 1\;P_1+ \pi\pi$   & $\frac{1}{3}A_0(1,1)+\frac{1}{12}A_1(1,1)+\frac{1}{12}A_2(1,1)~^a$ & 0.0003  \\
$2\;P_1\rightarrow 1^3P_0+ \pi\pi$   & $\frac{1}{9}A_1(1,1)$ & 0 \\
\hline
$2\;P'_1\rightarrow 1\;P_1+ \pi\pi$   & $\frac{1}{3}A_0(1,1)+\frac{1}{3}A_1(1,1)+\frac{1}{3}A_2(1,1)~^b$ & $3\times 10^{-6}$ \\
$2\;P'_1\rightarrow 1\;P'_1+ \pi\pi$  & $\frac{1}{3}A_0(1,1)+\frac{1}{3}A_1(1,1)+\frac{1}{3}A_2(1,1)~^b$ & 0.0014 \\
$2\;P'_1\rightarrow 1^3P_0+ \pi\pi$   & $\frac{1}{9}A_1(1,1)$ & 0 \\
\hline\hline
\end{tabular}\\
\begin{flushleft}
\hspace{2cm}$^a$The expression is for $^3P_1\rightarrow~^3P_1$ transition.\\
\hspace{2cm}$^b$The expression is for $^1P_1\rightarrow~^1P_1$ transition.
\end{flushleft}
\label{table:haronic-rates-1}
\end{table}

\begin{table}[H]
\centering
\caption{Rates for the E1-E1 hadronic transitions of 1D and 1F states of $B_c$ mesons.}
\begin{tabular}{c c c}
\hline\hline
Transition\hspace{.1 in}&\hspace{.1 in} Expression for the rate \hspace{.1 in}&\hspace{.1 in} the $c\overline{b}$ rate (keV) \hspace{.1 in}\\
\hline
$1^3D_1\rightarrow 1^3S_1+ \pi\pi$& $\frac{1}{5}A_2(2,0)$ & 0.042 \\
$1^3D_3\rightarrow 1^3S_1+ \pi\pi$& $\frac{1}{5}A_2(2,0)$ & 0.043 \\
$1\;D_2\rightarrow 1^1S_0+ \pi\pi$    & $\frac{1}{5}A_2(2,0)$ &  0.035\\
$1\;D_2\rightarrow 1^3S_1+ \pi\pi$    & $\frac{1}{5}A_2(2,0)$ &  0.017\\
$1\;D'_2\rightarrow 1^1S_0+ \pi\pi$   & $\frac{1}{5}A_2(2,0)$ &  0.23\\
$1\;D'_2\rightarrow 1^3S_1+ \pi\pi$   & $\frac{1}{5}A_2(2,0)$ &  0.026\\
\hline
$1^3F_2\rightarrow 1^3P_0+ \pi\pi$& $\frac{1}{15}A_2(3,1)$ & 0.0004 \\
$1^3F_2\rightarrow 1^3P_2+ \pi\pi$& $\frac{1}{105}A_2(3,1)$ & 0.00004 \\
$1^3F_2\rightarrow 1\;P_1+ \pi\pi$& $\frac{1}{15}A_2(3,1)$ & 0.0001 \\
$1^3F_2\rightarrow 1\;P_1'+ \pi\pi$& $\frac{1}{15}A_2(3,1)$ & 0.0002 \\
\hline
$1\;F_3\rightarrow 1^3P_2+ \pi\pi$& $\frac{1}{21}A_2(3,1)$ & 0.0001 \\
$1\;F_3\rightarrow 1\;P_1+ \pi\pi$& $\frac{2}{21}A_2(3,1)~^a$ & 0.001 \\
$1\;F_3\rightarrow 1\;P'_1+ \pi\pi$& $\frac{2}{21}A_2(3,1)~^a$ & $\sim 0$ \\
\hline
$1\;F'_3\rightarrow 1^3P_2+ \pi\pi$& $\frac{1}{21}A_2(3,1)$ & 0.0001 \\
$1\;F'_3\rightarrow 1\;P_1+ \pi\pi$& $\frac{1}{7}A_2(3,1)~^b$ & 0.00002 \\
$1\;F'_3\rightarrow 1\;P'_1+ \pi\pi$& $\frac{1}{7}A_2(3,1)~^b$ & 0.001 \\
\hline
$1^3F_4\rightarrow 1^3P_2+ \pi\pi$& $\frac{1}{7}A_2(3,1)$ & 0.0005 \\
\hline\hline
\end{tabular}\\
\begin{flushleft}
\hspace{3cm}$^a$The expression is for $^3F_3\rightarrow~^3P_1$ transition.\\
\hspace{3cm}$^b$The expression is for $^1F_3\rightarrow~^1P_1$ transition.
\end{flushleft}
\label{table:haronic-rates-2}
\end{table}

In Tables \ref{results-1} - \ref{results-8}, we combine the widths of radiative decays, strong decays, and hadronic transitions to calculate the total widths and the branching ratios. These BR's are used in the next section to give estimates for the number of events expected at the LHC for different decay chains of $B_c$ states below the threshold. In these tables $c_P=\cos\phi_{nP}$, $s_P=\sin\phi_{nP}$, $c_D=\cos\phi_{nD}$, $s_D=\sin\phi_{nD}$, $c_F=\cos\phi_{nF}$, and $s_F=\sin\phi_{nF}$, with $n$ being the principal quantum number.

\begin{table}[h!]
\tabcolsep=1pt\fontsize{9}{9}\selectfont
\centering
\renewcommand{\arraystretch}{0.8}
\caption{Partial widths and branching ratios for strong, radiative and hadronic transitions for the 1S, 2S, 3S and 4S states of $B_c$ mesons. Column 4, labeled $\mathcal{M}$ gives the matrix element appropriate to the particular decay. For E1 and M1 transitions matrix elements are $\langle \psi_f|r|\psi_i\rangle$ and $\langle \psi_f|j_0(kr\frac{m_{b,c}}{m_c+m_b})|\psi_i\rangle$ are in units of $\text{GeV}^{-1}$ and the strong decay amplitudes are in units of $\text{GeV}^{-1/2}$. Details of the calculations are given in the text.}
\begin{tabular}{c c c c c c}
\hline\hline
Meson\hspace{.1 in}&\hspace{.1 in} Decay Mode \hspace{.1 in}&\hspace{.1 in} Photon Energy\hspace{.1 in} & Amplitude($\mathcal{M}$) &\hspace{.1 in} $\Gamma_{thy}$\hspace{.1 in} &\hspace{.1 in} B.R \hspace{.1 in}\\
      &       & MeV       &              & MeV                        &($\%$) \\
\hline
$B^*_c$         & $B_c\gamma$       & 60.81  & $\langle 1^1S_0|j_0(kr\frac{m_{b,c}}{m_c+m_b})|1^3S_1\rangle=0.9983,0.999$ & 0.15~keV & 100\\
\hline
$B_c(2 ^1S_0)$  &$B_c(1 ^1S_0)+\pi\pi$ &     &               & 10.68~keV & 51.42\\
                &$B_c^*\gamma$      & 392.83 &$\langle 1^3S_1|j_0(kr\frac{m_{b,c}}{m_c+m_b})|2^1S_0\rangle=0.0316,-0.0182$ & 0.07~keV & 0.34\\
                &$B_c(1P_1)\gamma$  & 90.39  &$\langle 1^1P_1|r|2^1S_0\rangle=-2.8218$        & 7.11~keV & 34.23\\
                &$B_c(1P'_1)\gamma$ & 84.46  &$\langle 1^1P_1|r|2^1S_0\rangle=-2.8218$        & 2.91~keV & 14.01\\
                &total              &&                       & 20.77~keV & 100 \\
\hline
$B_c(2 ^3S_1)$  &$B_c(1 ^3S_1)+\pi\pi$ &     &               & 3.99~keV & 26.76\\
                &$B_c\gamma$        & 455.58 &$\langle 1^1S_0|j_0(kr\frac{m_{b,c}}{m_c+m_b})|2^3S_1\rangle=0.0887,0.0303$  & 0.41~keV & 2.75\\
                &$B_c(2^1S_0)\gamma$& 6.0    &$\langle 2^1S_0|j_0(kr\frac{m_{b,c}}{m_c+m_b})|2^3S_1\rangle=0.9994,0.9995$ & $1.4\times 10^{-4}$~keV &0.001\\
                &$B_c(1^3P_2)\gamma$& 81.5   &$\langle 1^3P_2|r|2^3S_1\rangle=-2.8099$        & 4.31~keV & 28.91\\
                &$B_c(1P_1)\gamma$  & 96.3   &$\langle 1^3P_1|r|2^3S_1\rangle=-2.8099$        & 1.43~keV & 9.59 \\
                &$B_c(1P'_1)\gamma$ & 90.39  &$\langle 1^3P_1|r|2^3S_1\rangle=-2.8099$        & 2.35~keV & 15.76\\
                &$B_c(1^3P_0)\gamma$& 115.0  &$\langle 1^3P_0|r|2^3S_1\rangle=-2.8099$        & 2.42~keV & 16.23\\
                &total              &&                       & 14.91~keV & 100 \\
\hline
$B_c(3 ^1S_0)$ &$B_c(1 ^1S_0)+\pi\pi$ &     &               & 0.97~keV & 4.71\\
               &$B_c^*\gamma$       & 645.23 &$\langle 1^3S_1|j_0(kr\frac{m_{b,c}}{m_c+m_b})|3^1S_0\rangle=0.0259,-0.0079$ & 0.24~keV & 1.17\\
               &$B_c(2^3S_1)\gamma$ & 261.92 &$\langle 2^3S_1|j_0(kr\frac{m_{b,c}}{m_c+m_b})|3^1S_0\rangle=0.0546,-0.0101$ & 0.07~keV & 0.34\\
               &$B_c(1P_1)\gamma$   & 354.55 & $\langle 1^3P_1|r|3^1S_0\rangle=0.1055$        & 0.6~keV  & 2.92\\
               &$B_c(1P'_1)\gamma$  & 348.86 & $\langle 1^3P_1|r|3^1S_0\rangle=0.1055$        & 0.29~keV & 1.41\\
               &$B_c(2P_1)\gamma$   & 83.50  & $\langle 2^3P_1|r|3^1S_0\rangle=-4.3884$       & 13.56~keV & 65.89\\
               &$B_c(2P'_1)\gamma$  & 74.60  & $\langle 2^3P_1|r|3^1S_0\rangle=-4.3884$       & 4.85~keV & 23.57\\
               & total              &&                       & 20.58~keV & 100 \\
\hline
$B_c(3^3S_1)$  &$B_c(1 ^3S_1)+\pi\pi$ &     &               & 0.81~keV & 4.12\\
              &$B_c\gamma$           & 703.76 & $\langle 1^1S_0|j_0(kr\frac{m_{b,c}}{m_c+m_b})|3^3S_1\rangle=0.0552,0.0145$ & 0.57~keV & 2.9\\
             &$B_c(2^1S_0)\gamma$   & 271.53 & $\langle 2^1S_0|j_0(kr\frac{m_{b,c}}{m_c+m_b})|3^3S_1\rangle=0.0889,0.0246$ & 0.08~keV & 0.41\\
             &$B_c(3^1S_0)\gamma$   & 4.0    & $\langle 3^1S_0|j_0(kr\frac{m_{b,c}}{m_c+m_b})|3^3S_1\rangle=0.9996,0.9996$ & $4.2\times 10^{-5}$~keV &0.0002\\
             &$B_c(1^3P_2)\gamma$   & 344.12 & $\langle 1^3P_2|r|3^3S_1\rangle=0.0926$        & 0.35~keV & 1.78\\
             &$B_c(1P_1)\gamma$     & 358.35 & $\langle 1^3P_1|r|3^3S_1\rangle=0.0926$        & 0.08~keV & 0.41\\
             &$B_c(1P'_1)\gamma$    & 352.66 & $\langle 1^3P_1|r|3^3S_1\rangle=0.0926$        & 0.15~keV & 0.76\\
             &$B_c(1^3P_0)\gamma$   & 376.33 & $\langle 1^3P_0|r|3^3S_1\rangle=0.0926$        & 0.09~keV & 0.46\\
             &$B_c(2^3P_2)\gamma$   & 71.63  & $\langle 2^3P_2|r|3^3S_1\rangle=-4.3738$       & 7.09~keV & 36.03\\
             &$B_c(2P_1)\gamma$     & 87.45  & $\langle 2^3P_1|r|3^3S_1\rangle=-4.3738$       & 2.59~keV & 13.16\\
             &$B_c(2P'_1)\gamma$    & 78.56  & $\langle 2^3P_1|r|3^3S_1\rangle=-4.3738$       & 3.74~keV & 19.0\\
             &$B_c(2^3P_0)\gamma$   & 102.24 & $\langle 2^3P_0|r|3^3S_1\rangle=-4.3738$       & 4.13~keV & 20.99\\
             &total                 &&                       & 19.68~keV & 100 \\
\hline
$B_c(4 ^1S_0)$ &$D B^*$             && $^3P_0=+0.0305$        & 2.96  & 100\\
\hline
$B_c(4 ^3S_1)$ &$D B$               &        & $^1P_1=-0.0067$                                                & 0.21 & 11.66 \\
               &$D B^*$             && $^3P_1=-0.022 $                                                & 1.59 & 88.29\\
               &total               &&                                                                & 1.8     &100 \\
\hline\hline
\end{tabular}
\label{results-1}
\end{table}

\begin{table}[h!]
\tabcolsep=1pt\fontsize{10}{10}\selectfont
\centering
\renewcommand{\arraystretch}{0.8}
\caption{Partial widths and branching ratios for strong, radiative and hadronic transitions for the 1P, 2P and 3P states of $B_c$ mesons (format as in the Table~\ref{results-1}).}
\begin{tabular}{c c c c c c}
\hline\hline
Meson\hspace{.1 in}&\hspace{.1 in} Decay Mode \hspace{.1 in}&\hspace{.1 in} Photon Energy\hspace{.1 in} & Amplitude($\mathcal{M}$) &\hspace{.1 in} $\Gamma_{thy}$\hspace{.1 in} &\hspace{.1 in} B.R \hspace{.1 in}\\
      &       & MeV       &              & MeV                        &($\%$) \\
\hline
$B_c(1 ^3P_0)$      & $B^*_c\gamma$      & 288.44 & $\langle 1^3S_1|r|1^3P_0\rangle=1.896$ & 52.23~keV & 100\\
\hline
$B_c(1\;P_1)$       & $B_c\gamma$        & 323.71 & $\langle 1^1S_0|r|1^1P_1\rangle=1.8362$ & 46.13~keV & 68.78\\
                    & $B_c^*\gamma$      & 306.59 & $\langle 1^3S_1|r|1^3P_1\rangle=1.896$  & 20.94~keV & 31.22\\
                    &  total             &&                & 67.07~keV & 100\\
\hline
$B_c(1\;P'_1)$      & $B_c\gamma$        & 329.42 &$\langle 1^1S_0|r|1^1P_1\rangle=1.8362$ & 24.37~keV & 35.56\\
                    & $B_c^*\gamma$      & 312.31 & $\langle 1^3S_1|r|1^3P_1\rangle=1.896$  & 44.16~keV & 64.44\\
                    &  total             &&                & 68.53~keV & 100\\
\hline
$B_c(1 ^3P_2)$      & $B^*_c\gamma$      & 320.88 & $\langle 1^3S_1|r|1^3P_2\rangle=1.896$  & 71.91~keV & 100\\
\hline
$B_c(2 ^3P_0)$      &$B_c(1 ^3P_0)+\pi\pi$&       &        &  0.002~keV &  0.004         \\
                    &$B^*_c\gamma$       & 554.76 & $\langle 1^3S_1|r|2^3P_0\rangle=0.4117$ & 17.52~keV & 35.87\\
                    &$B_c(2^3S_1)\gamma$ & 165.96 & $\langle 2^3S_1|r|2^3P_0\rangle=3.1266$ & 27.05~keV & 55.38\\
                    &$B_c(1^3D_1)\gamma$ & 73.60  &$\langle 1^3D_1|r|2^3P_0\rangle=-2.9727$ & 4.27~keV  & 8.74\\
                    &total               &&                & 48.84~keV & 100\\
\hline
$B_c(2\;P_1)$       &$B_c(1\;P_1)+\pi\pi$&        &        & 0.0003~keV  & 0.0005          \\
                    &$B_c\gamma$         & 584.98 & $\langle 1^1S_0|r|2^1P_1\rangle=0.4258$ & 14.64~keV & 23.16\\
                    &$B^*_c\gamma$       & 568.54 & $\langle 1^3S_1|r|2^3P_1\rangle=0.4117$ & 6.3~keV   & 9.97\\
                    &$B_c(2^1S_0)\gamma$ & 186.42 & $\langle 2^1S_0|r|2^1P_1\rangle=3.0728$ & 24.67~keV & 39.02\\
                    &$B_c(2^3S_1)\gamma$ & 180.58 & $\langle 2^3S_1|r|2^3P_1\rangle=3.1266$ & 11.64~keV & 18.41\\
                    &$B_c(1\;D_2)\gamma$ & 84.48  & $\langle 1^3D_2|r|2^3P_1\rangle=-2.9727$,& 5.86~keV  & 9.27\\
                    &                    &        & $\langle 1^1D_2|r|2^1P_1\rangle=-2.9929$ \\
                    &$B_c(1\;D'_2)\gamma$ &84.48  & $\langle 1^3D_2|r|2^3P_1\rangle=-2.9727$,& 0.11~keV  & 0.17\\
                    &                    &        & $\langle 1^1D_2|r|2^1P_1\rangle=-2.9929$ \\
                    &total               &&                & 63.22~keV & 100\\
\hline
$B_c(2\;P'_1)$      &$B_c(1\;P_1)+\pi\pi$&        &        & $3\times 10^{-6}$~keV  & $\sim 0$         \\
                    &$B_c(1\;P'_1)+\pi\pi$&        &        & 0.0014~keV  & 0.002          \\
                    &$B_c\gamma$         & 593.21 & $\langle 1^1S_0|r|2^1P_1\rangle=0.4258$ & 7.65~keV  & 10.83\\
                    &$B^*_c\gamma$       & 576.80 & $\langle 1^3S_1|r|2^3P_1\rangle=0.4117$ & 13.12~keV & 18.57\\
                    &$B_c(2^1S_0)\gamma$ & 195.18 & $\langle 2^1S_0|r|2^1P_1\rangle=3.0728$ & 14.19~keV & 20.08\\
                    &$B_c(2^3S_1)\gamma$ & 189.34 & $\langle 2^3S_1|r|2^3P_1\rangle=3.1266$ & 26.76~keV & 37.87\\
                    &$B_c(1^3D_1)\gamma$ & 97.31  &$\langle 1^3D_1|r|2^3P_1\rangle=-2.9727$ &  1.64~keV &2.32 \\
                    &$B_c(1\;D_2)\gamma$ & 93.36  & $\langle 1^3D_2|r|2^3P_1\rangle=-2.9727$,& $1.5\times 10^{-4}$~keV  & 0.0002 \\
                    &                    &        & $\langle 1^1D_2|r|2^1P_1\rangle=-2.9929$ \\
                    &$B_c(1\;D'_2)\gamma$ & 93.36 & $\langle 1^3D_2|r|2^3P_1\rangle=-2.9727$,& 7.3~keV  & 10.33\\
                    &                    &        & $\langle 1^1D_2|r|2^1P_1\rangle=-2.9929$ \\
                    &total               &&                & 70.66~keV & 100\\
\hline
$B_c(2 ^3P_2)$      &$B_c(1^3P_2)+\pi\pi$&        &        & 0.0007~keV  & 0.0009          \\
                    &$B_c(1\;P_1)+\pi\pi$&        &        & 0.0002~keV  & 0.0003          \\
                    &$B_c(1\;P'_1)+\pi\pi$&        &        & 0.0002~keV  & 0.0003          \\
                    &$B_c(1^3P_0)+\pi\pi$&        &        & 0.0005~keV  &   0.0007        \\
                    &$B^*_c\gamma$       & 583.21 &$\langle 1^3S_1|r|2^3P_2\rangle=0.4117$  & 20.36~keV & 27.03\\
                    &$B_c(2^3S_1)\gamma$ & 196.15 &$\langle 2^3S_1|r|2^3P_2\rangle=3.1266$  & 44.67~keV &  59.31\\
                    &$B_c(1^3D_1)\gamma$ & 104.21 &$\langle 1^3D_1|r|2^3P_2\rangle=-2.9727$ & 0.12~keV  & 0.16\\
                    &$B_c(1^3D_3)\gamma$ & 98.29  &$\langle 1^3D_3|r|2^3P_2\rangle=-2.9727$ & 8.54~keV  & 11.34\\
                    &$B_c(1\;D_2)\gamma$ & 100.27 &$\langle 1^3D_2|r|2^3P_2\rangle=-2.9727$ & 0.65~keV  & 0.86\\
                    &$B_c(1\;D'_2)\gamma$ & 100.27&$\langle 1^3D_2|r|2^3P_2\rangle=-2.9727$ & 0.97~keV  & 1.29\\
                    &total               &&                & 75.31~keV & 100          \\
\hline
$B_c(3 ^3P_0)$      &$D B$      && $^1S_0=+0.1157$ & 3.34  & 100 \\
\hline
$B_c(3 ^3P_2)$      &$D B$      && $^1D_2=-0.0215$ & 1.16  & 100\\
\hline\hline
\end{tabular}
\label{results-2}
\end{table}

\begin{table}[h!]
\tabcolsep=1pt\fontsize{10}{10}\selectfont
\centering
\renewcommand{\arraystretch}{0.8}
\caption{Partial widths and branching ratios for strong, radiative and hadronic transitions for the 4P states of $B_c$ mesons (format as in the Table~\ref{results-1}).}
\begin{tabular}{c c c c c c}
\hline\hline
Meson\hspace{.1 in}&\hspace{.1 in} Decay Mode \hspace{.1 in}&\hspace{.1 in} Photon Energy\hspace{.1 in} & Amplitude($\mathcal{M}$) &\hspace{.1 in} $\Gamma_{thy}$\hspace{.1 in} &\hspace{.1 in} B.R \hspace{.1 in}\\
      &       & MeV       &              & MeV                        &($\%$) \\
\hline
$B_c(4 ^3P_0)$     &$D B$      && $^1S_0=+0.008$           & 0.45 & 9.03 \\
                   &$D^* B^*$  && $^1S_0=-0.0196$          & 4.53 & 90.93\\
                   &           && $^5D_0=-0.0429$          &      & \\
                   &$D_s B_s$  && $^1S_0=+0.001$           & 0.002& 0.04\\
                   &total      &&                          & 4.98     & 100\\
\hline
$B_c(4\;P_1)$      &$D B^*$    && $^3S_1=-0.0083c_P+0.0117s_P$ & 1.03 & 14.21\\
                   &           && $^3D_1=-0.0104c_P-0.0072s_P$ &      & \\
                   &$D^* B$    && $^3S_1=-0.0127c_P+0.0176s_P$ & 0.94 & 12.97\\
                   &           && $^3D_1=-0.0120c_P-0.0082s_P$ &      &\\
                   &$D^* B^*$  && $^3S_1=+0.0006c_P$,        & 5.28 & 72.83\\
                   &           && $^3D_1=+0.0406c_P$,        &      &\\
                   &           && $^5D_1=-0.0491s_P$         &      &\\
                   &total      &&                          & 7.25     & 100\\
\hline
$B_c(4\;P'_1)$     &$D B^*$    && $^3S_1=+0.0120c_P+0.0086s_P$ & 1.44 & 13.81\\
                   &           && $^3D_1=-0.0069c_P+0.0100s_P$ &      & \\
                   &$D^* B$    && $^3S_1=+0.0153c_P+0.0111s_P$ & 1.67 & 16.01\\
                   &           && $^3D_1=-0.0053c_P+0.0079s_P$ &      & \\
                   &$D^* B^*$  && $^3S_1=-0.0121s_P$,        & 7.32 & 70.18\\
                   &           && $^3D_1=-0.0417s_P$,        &      &\\
                   &           && $^5D_1=-0.0503c_P$         &      &\\
                   &total      &&                          &  10.43    & 100\\
\hline
$B_c(4 ^3P_2)$     &$D B$      && $^1D_2=-0.0013$ & 0.01 & 0.14\\
                   &$D B^*$    && $^3D_2=+0.0088$ & 0.52 & 7.38\\
                   &$D^* B$    && $^3D_2=+0.0001$ & 0.1  & 1.42\\
                   &$D^* B^*$  && $^5S_2=+0.0246$ & 6.41 & 91.01\\
                   &           && $^1D_2=+0.0126$ &      &\\
                   &           && $^5D_2=-0.0335$ &      &\\
                   &$D_s B_s$  && $^1D_2=-0.001$  & 0.003& 0.004 \\
                   &total      &&                 &  7.043    & 100 \\
\hline\hline
\end{tabular}
\label{results-3}
\end{table}

\begin{table}[h!]
\tabcolsep=1pt\fontsize{10}{10}\selectfont
\centering
\renewcommand{\arraystretch}{0.8}
\caption{Partial widths and branching ratios for strong, radiative and hadronic transitions for the 1D states of $B_c$ mesons (format as in the Table~\ref{results-1}).}
\begin{tabular}{c c c c c c}
\hline\hline
Meson\hspace{.1 in}&\hspace{.1 in} Decay Mode \hspace{.1 in}&\hspace{.1 in} Photon Energy\hspace{.1 in} & Amplitude &\hspace{.1 in} $\Gamma_{thy}$\hspace{.1 in} &\hspace{.1 in} B.R \hspace{.1 in}\\
      &       & MeV       &              & MeV                        &($\%$) \\
\hline
$B_c(1 ^3D_1)$    & $B_c(1^3S_1)+\pi\pi$ &  &  & 0.042~keV & 0.07 \\
                  & $B_c(1^3P_0)\gamma$ & 206.78 & $\langle 1^3P_0|r|1^3D_1\rangle=3.3708$ & 40.55~keV & 63.97 \\
                  & $B_c(1P_1)\gamma$   & 188.33 & $\langle 1^3P_1|r|1^3D_1\rangle=3.3708$ & 7.67~keV & 12.1 \\
                  & $B_c(1P'_1)\gamma$  & 182.50 & $\langle 1^3P_1|r|1^3D_1\rangle=3.3708$ & 13.93~keV &21.98 \\
                  & $B_c(1^3P_2)\gamma$ & 173.74 & $\langle 1^3P_2|r|1^3D_1\rangle=3.3708$ & 1.2~keV & 1.89\\
                  &total                &&                & 63.39~keV & 100   \\
\hline
$B_c(1\;D_2)$     & $B_c(1^1S_0)+\pi\pi$ &  &  & 0.035~keV &  0.06\\
                  & $B_c(1^3S_1)+\pi\pi$ &  &  & 0.017~keV &  0.03\\
                  & $B_c(1^3P_2)\gamma$ & 177.63 & $\langle 1^3P_2|r|1^3D_2\rangle=3.3708$ & 4.62~keV & 8.12   \\
                  & $B_c(1\;P_1)\gamma$ & 192.22 & $\langle 1^3P_1|r|1^3D_2\rangle=3.3708$,& 52.26~keV & 91.8    \\
                  &                     &        & $\langle 1^1P_1|r|1^1D_2\rangle=3.3461$ \\
                  & $B_c(1\;P'_1)\gamma$ &186.39 & $\langle 1^3P_1|r|1^3D_2\rangle=3.3708$, & $2.6\times10^{-4}$~keV & 0.0005   \\
                  &                     &        & $\langle 1^1P_1|r|1^1D_2\rangle=3.3461$ \\
                  &total                &&                & 56.93~keV & 100   \\
\hline
$B_c(1\;D'_2)$    & $B_c(1^1S_0)+\pi\pi$ &  &  & 0.023~keV &  0.04\\
                  & $B_c(1^3S_1)+\pi\pi$ &  &  & 0.026~keV &  0.05\\
                  & $B_c(1^3P_2)\gamma$ & 177.63 & $\langle 1^3P_2|r|1^3D_2\rangle=3.3708$ & 6.95~keV & 13.32  \\
                  & $B_c(1\;P_1)\gamma$ & 192.22 & $\langle 1^3P_1|r|1^3D_2\rangle=3.3708$ & 0.9~keV & 1.72   \\
                  &                     &        & $\langle 1^1P_1|r|1^1D_2\rangle=3.3461$ \\
                  & $B_c(1\;P'_1)\gamma$ &186.39 & $\langle 1^3P_1|r|1^3D_2\rangle=3.3708$ & 44.29~keV &  84.86  \\
                  &                     &        & $\langle 1^1P_1|r|1^1D_2\rangle=3.3461$ \\
                  &total                &&                & 52.19~keV & 100   \\
\hline
$B_c(1 ^3D_3)$    & $B_c(1^3S_1)+\pi\pi$ &  &  & 0.043~keV & 0.09 \\
                  & $B_c(1^3P_2)\gamma$ & 179.58 & $\langle 1^3P_2|r|1^3D_3\rangle=3.3708$ & 47.81~keV & 99.91  \\
                  &total                &&                & 47.85~keV & 100   \\
\hline\hline
\end{tabular}
\label{results-4}
\end{table}

\begin{table}[h!]
\tabcolsep=1pt\fontsize{10}{10}\selectfont
\centering
\renewcommand{\arraystretch}{0.8}
\caption{Partial widths and branching ratios for strong, radiative and hadronic transitions for the 2D states of $B_c$ mesons (format as in the Table~\ref{results-1}).}
\begin{tabular}{c c c c c c}
\hline\hline
Meson\hspace{.1 in}&\hspace{.1 in} Decay Mode \hspace{.1 in}&\hspace{.1 in} Photon Energy\hspace{.1 in} & Amplitude &\hspace{.1 in} $\Gamma_{thy}$\hspace{.1 in} &\hspace{.1 in} B.R \hspace{.1 in}\\
      &       & MeV       &              & MeV                        &($\%$) \\
\hline
$B_c(2 ^3D_1)$    & $B_c(2^3P_0)\gamma$ & 163.08 & $\langle 2^3P_0|r|2^3D_1\rangle=4.4775$ & 35.09~keV &52.98 \\
                  & $B_c(2P_1)\gamma$   & 148.41 & $\langle 2^3P_1|r|2^3D_1\rangle=4.4775$ & 6.62~keV  &10.0 \\
                  & $B_c(2P'_1)\gamma$  & 139.60 & $\langle 2^3P_1|r|2^3D_1\rangle=4.4775$ & 11.0~keV  & 16.61\\
                  & $B_c(2^3P_2)\gamma$ & 132.73 & $\langle 2^3P_2|r|2^3D_1\rangle=4.4775$ & 0.95~keV  & 1.43\\
                  & $B_c(1^3P_0)\gamma$ & 434.76 & $\langle 1^3P_0|r|2^3D_1\rangle=0.3469$ & 3.99~keV  & 6.02\\
                  & $B_c(1P_1)\gamma$   & 416.94 & $\langle 1^3P_1|r|2^3D_1\rangle=0.3469$ & 0.88~keV  & 1.33\\
                  & $B_c(1P'_1)\gamma$  & 411.30 & $\langle 1^3P_1|r|2^3D_1\rangle=0.3469$ & 1.69~keV  & 2.55\\
                  & $B_c(1^3P_2)\gamma$ & 402.84 & $\langle 1^3P_2|r|2^3D_1\rangle=0.3469$ & 0.16~keV  & 0.24\\
                  & $B_c(1^3F_2)\gamma$ & 82.32  & $\langle 1^3F_2|r|2^3D_1\rangle=-3.1023$& 5.85~keV  &  8.83\\
                  &total                &&                & 66.23~keV & 100   \\
\hline
$B_c(2\;D_2)$     & $B_c(2^3P_2)\gamma$ & 136.66 & $\langle 2^3P_2|r|2^3D_2\rangle=4.4775$ & 3.71~keV & 5.82   \\
                  & $B_c(2\;P_1)\gamma$ & 152.33 & $\langle 2^3P_1|r|2^3D_2\rangle=4.4775$,& 45.81~keV & 71.84   \\
                  &                     &  & $\langle 2^1P_1|r|2^1D_2\rangle=4.4392$ & \\
                  & $B_c(2\;P'_1)\gamma$& 143.52 & $\langle 2^3P_1|r|2^3D_2\rangle=4.4775$,& $3.2\times 10^{-4}$~keV & 0.001   \\
                  &                     &  & $\langle 2^1P_1|r|2^1D_2\rangle=4.4392$ & \\
                  & $B_c(1^3P_2)\gamma$ & 406.61 & $\langle 1^3P_2|r|2^3D_2\rangle=0.3469$ & 0.59~keV &    0.93\\
                  & $B_c(1\;P_1)\gamma$ & 420.71 & $\langle 1^3P_1|r|2^3D_2\rangle=0.3469$, & 6.17~keV &   9.68 \\
                  &                     &  & $\langle 1^1P_1|r|2^1D_2\rangle=0.3603$ & \\
                  & $B_c(1\;P'_1)\gamma$& 415.07 & $\langle 1^3P_1|r|2^3D_2\rangle=0.3469$, & $2.1\times 10^{-3}$~keV & 0.003   \\
                  &                     &  & $\langle 1^1P_1|r|2^1D_2\rangle=0.3603$ & \\
                  & $B_c(1^3F_2)\gamma$ & 86.27 & $\langle 1^3F_2|r|2^3D_2\rangle=-3.1023$ & 0.3~keV &  0.47  \\
                  & $B_c(1\;F_3)\gamma$ & 89.43 & $\langle 1^3F_3|r|2^3D_2\rangle=-3.1023$, & 7.16~keV & 11.23   \\
                  &                     &  & $\langle 1^1F_3|r|2^1D_2\rangle=-3.1085$ & \\
                  & $B_c(1\;F'_3)\gamma$& 82.51 & $\langle 1^3F_3|r|2^3D_2\rangle=-3.1023$, & 0.02~keV & 0.03   \\
                  &                     &  & $\langle1^1F_3|r|2^1D_2\rangle=-3.1085$ & \\
                  &total                &&                & 63.76~keV & 100   \\
\hline
$B_c(2\;D'_2)$    & $B_c(2^3P_2)\gamma$ & 136.66 & $\langle 2^3P_2|r|2^3D_2\rangle=4.4775$ & 5.58~keV & 10.26   \\
                  & $B_c(2\;P_1)\gamma$ & 152.33 & $\langle 2^3P_1|r|2^3D_2\rangle=4.4775$, & 0.79~keV & 1.45   \\
                  &                     &  & $\langle 2^1P_1|r|2^1D_2\rangle=4.4392$  & \\
                  & $B_c(2\;P'_1)\gamma$& 143.52 & $\langle 2^3P_1|r|2^3D_2\rangle=4.4775$, & 35.64~keV & 65.56   \\
                  &                     &  & $\langle 2^1P_1|r|2^1D_2\rangle=4.4392$  & \\
                  & $B_c(1^3P_2)\gamma$ & 406.61 & $\langle 1^3P_2|r|2^3D_2\rangle=0.3469$ & 0.88~keV & 1.62   \\
                  & $B_c(1\;P_1)\gamma$ & 420.71 & $\langle 1^3P_1|r|2^3D_2\rangle=0.3469$, & 0.14~keV & 0.26   \\
                  &                     &  & $\langle 1^1P_1|r|2^1D_2\rangle=0.3603$ \\
                  & $B_c(1\;P'_1)\gamma$& 415.07 & $\langle 1^3P_1|r|2^3D_2\rangle=0.3469$, & 5.37~keV & 9.88   \\
                  &                     &  & $\langle 1^1P_1|r|2^1D_2\rangle=0.3603$ \\
                  & $B_c(1^3F_2)\gamma$ & 86.27  & $\langle 1^3F_2|r|2^3D_2\rangle=-3.1023$ & 0.45~keV & 0.83   \\
                  & $B_c(1\;F_3)\gamma$ & 89.43 & $\langle 1^3F_2|r|2^3D_2\rangle=-3.1023$, & $1.7\times 10^{-6}$~keV & $\sim 0$   \\
                  &                     &  & $\langle 1^1F_2|r|2^1D_2\rangle=-3.1085$ \\
                  & $B_c(1\;F'_3)\gamma$& 82.51 & $\langle 1^3F_2|r|2^3D_2\rangle=-3.1023$, & 5.51~keV & 10.14   \\
                  &                     &  & $\langle 1^1F_2|r|2^1D_2\rangle=-3.1085$ \\
                  &total                &&                &  54.36~keV & 100   \\
\hline
$B_c(2 ^3D_3)$    & $B_c(2^3P_2)\gamma$ & 139.60 & $\langle 2^3P_2|r|2^3D_3\rangle=4.4775$ & 39.62~keV & 74.5   \\
                  & $B_c(1^3P_2)\gamma$ & 409.44 & $\langle 1^3P_2|r|2^3D_3\rangle=0.3469$ & 6.0~keV & 11.28   \\
                  & $B_c(1^3F_4)\gamma$ & 89.72  & $\langle 1^3F_4|r|2^3D_3\rangle=-3.1023$& 6.96~keV & 13.09   \\
                  & $B_c(1^3F_2)\gamma$ & 89.23  & $\langle 1^3F_2|r|2^3D_3\rangle=-3.1023$& 0.02~keV & 0.04   \\
                  & $B_c(1F_3)\gamma$   & 92.39  & $\langle 1^3F_3|r|2^3D_3\rangle=-3.1023$& 0.28~keV & 0.53   \\
                  & $B_c(1F'_3)\gamma$  & 85.48  & $\langle 1^3F_3|r|2^3D_3\rangle=-3.1023$& 0.3~keV  & 0.56   \\
                  &total                &&                                   & 53.18~keV & 100   \\
\hline\hline
\end{tabular}
\label{results-5}
\end{table}

\begin{table}[h!]
\tabcolsep=1pt\fontsize{10}{10}\selectfont
\centering
\renewcommand{\arraystretch}{0.8}
\caption{Partial widths and branching ratios for strong, radiative and hadronic transitions for the 3D states of $B_c$ mesons (format as in the Table~\ref{results-1}).}
\begin{tabular}{c c c c c c}
\hline\hline
Meson\hspace{.1 in}&\hspace{.1 in} Decay Mode \hspace{.1 in}&\hspace{.1 in} Photon Energy\hspace{.1 in} & Amplitude &\hspace{.1 in} $\Gamma_{thy}$\hspace{.1 in} &\hspace{.1 in} B.R \hspace{.1 in}\\
      &       & MeV       &              & MeV                        &($\%$) \\
\hline
$B_c(3 ^3D_1)$    &$D B$                && $^1P_1=-0.0126$                   & 0.91 & 59.48 \\
                  &$D B^*$              && $^3P_1=+0.0108$                   & 0.55 & 35.95 \\
                  &$D^* B$              && $^3P_1=+0.0122$                   & 0.07 & 4.58 \\
                  &total                &&                                   & 1.53     & 100\\
\hline
$B_c(3\;D_2)$     &$D B^*$              && $^3P_2=-0.0127c_D+0.0156s_D$      & 1.81 & 99.81  \\
                  &                     && $^3F_2=-0.0151c_D-0.0123s_D$      &      &  \\
                  &$D^* B$              && $^3P_2=-0.0281c_D+0.0344s_D$      & $3.5\times 10^{-3}$ & 0.19  \\
                  &                     && $^3F_2=-0.0014c_D-0.0011s_D$      &      & \\
                  &total                &&                                   &  1.81    & 100\\
\hline
$B_c(3\;D_2')$    &$D B^*$              && $^3D_2=+0.0156c_D+0.0127s_D$      & 1.93 & 47.3 \\
                  &                     && $^3F_2=-0.0123c_D+0.0151s_D$      &      &\\
                  &$D^* B$              && $^3D_2=+0.0344c_D+0.0281s_D$      & 2.15 & 52.7 \\
                  &                     && $^3F_2=-0.0011c_D+0.0014s_D$      &      &\\
                  &total                &&                                   & 4.08 & 100   \\
\hline
$B_c(3 ^3D_3)$    &$D B$                && $^1F_3=+0.0002$                   & 0.0002& 0.02   \\
                  &$D B^*$              && $^3F_3=-0.0143$                   & 0.99  & 98.98\\
                  &$D^* B$              && $^3F_3=-0.0027$                   & 0.01  & 1.0 \\
                  &total                &&                                   & 1.0      & 100  \\
\hline\hline
\end{tabular}
\label{results-6}
\end{table}

\begin{table}[h!]
\tabcolsep=1pt\fontsize{10}{10}\selectfont
\centering
\renewcommand{\arraystretch}{0.8}
\caption{Partial widths and branching ratios for strong, radiative and hadronic transitions for the 4D states of $B_c$ mesons (format as in the Table~\ref{results-1}).}
\begin{tabular}{c c c c c c}
\hline\hline
Meson\hspace{.1 in}&\hspace{.1 in} Decay Mode \hspace{.1 in}&\hspace{.1 in} Photon Energy\hspace{.1 in} & Amplitude($\mathcal{M}$) &\hspace{.1 in} $\Gamma_{thy}$\hspace{.1 in} &\hspace{.1 in} B.R \hspace{.1 in}\\
      &       & MeV       &              & MeV                        &($\%$) \\
\hline
$B_c(4 ^3D_1)$ &$D B$           && $^1P_1=-0.0178$  & 2.95  & 72.25\\
               &$D B^*$         && $^3P_1=+0.0023$  & 0.05  & 1.22\\
               &$D^* B$         && $^3P_1=+0.0061$  & 0.27  & 6.61\\
               &$D^* B^*$       && $^1P_1=+0.0009$  & 0.81  & 19.84\\
               &                && $^5P_1=-0.0004$  &       &\\
               &                && $^5F_1=-0.0114$  &       &\\
               &$D_s B_s$       && $^1P_1=-0.0007$  & 0.003 & 0.07     \\
               &$D_s B_s^*$     && $^3P_1=+0.0002$  & $1.5\times 10^{-4}$ & $\sim 0$         \\
               &total           &&                  &   4.08    & 100     \\
\hline
$B_c(4\;D_2)$  &$D B^*$         && $^3D_2=-0.0028c_D+0.0040s_D$            & 0.61 & 14.94  \\
               &                && $^3F_2=0.0063c_D+0.0055s_D$             &  \\
               &$D^* B$         && $^3D_2=-0.0072c_D+0.0089s_D$            & 1.03 & 25.23  \\
               &                && $^3F_2=0.0092c_D+0.0074s_D$             & \\
               &$D^* B^*$       && $^3P_2=-0.0005c_D$, $^3F_2=-0.0093c_D$  & 0.65 &15.92\\
               &                && $^5P_2=+0.0001s_D$, $^5F_2=+0.0109s_D$  &  \\
               &$D_s B_s^*$     && $^3D_2=-0.0001c_D+0.0001s_D$            & $7.0\times 10^{-4}$ & 0.02  \\
               &                && $^3F_2=+0.0003c_D+0.0003s_D$            & \\
               &$D_s^* B_s$     && $^3D_2=-0.0021c_D+0.0025s_D$            & $3.7\times 10^{-5}$ & $\sim 0$  \\
               &                && $^3F_2=+0.0001c_D+0.0001s_D$            &\\
               &total           &&                                         & 2.29 & 100\\
\hline
$B_c(4\;D_2')$ &$D B^*$         && $^3D_2=0.0040c_D+0.0028s_D$             & 0.21 & 11.28  \\
               &                && $^3F_2=0.0055c_D-0.0063s_D$             &      &\\
               &$D^* B$         && $^3D_2=0.0089c_D+0.0072s_D$             & 0.96 & 51.58  \\
               &                && $^3F_2=0.0074c_D-0.0092s_D$             &      &\\
               &$D^* B^*$       && $^3P_2=+0.0005s_D$, $^3F_2=+0.0093s_D$  & 0.68 & 36.54\\
               &                && $^5P_2=+0.0001c_D$, $^5F_2=+0.0109c_D$  &      &\\
               &$D_s B_s^*$     && $^3D_2=+0.0001c_D+0.0001s_D$ & $1.5\times 10^{-4}$ 0.01  \\
               &                && $^3F_2=+0.0003c_D-0.0003s_D$ \\
               &$D_s^* B_s$     && $^3D_2=+0.0025c_D+0.0021s_D$ & 0.011  & 0.59 \\
               &                && $^3F_2=+0.0001c_D-0.0001s_D$ \\
               &total           &&                                         & 1.86 & 100 \\
\hline
$B_c(4 ^3D_3)$ &$D B$           && $^1F_3=+0.0117$ & 1.3                 & 44.19\\
               &$D B^*$         && $^3F_3=-0.0074$ & 0.48                & 16.32\\
               &$D^* B$         && $^3F_3=-0.0088$ & 0.58                & 19.72\\
               &$D^* B^*$       && $^5P_3=-0.0024$ & 0.58                & 19.72\\
               &                && $^1F_3=+0.0038$ &                     &\\
               &                && $^5F_3=-0.0083$ &                     &\\
               &$D_s B_s$       && $^1F_3=+0.0003$ & $4.4\times 10^{-4}$ & 0.01     \\
               &$D_s B_s^*$     && $^3F_3=-0.0004$ & $8.9\times 10^{-4}$ & 0.03     \\
               &$D_s^* B_s$     && $^3F_3=-0.0005$ & $3.5\times 10^{-4}$ & 0.01      \\
               &total           &&                 &  2.94                   & 100         \\
\hline\hline
\end{tabular}
\label{results-7}
\end{table}

\begin{table}[h!]
\tabcolsep=1pt\fontsize{10}{10}\selectfont
\centering
\renewcommand{\arraystretch}{0.8}
\caption{Partial widths and branching ratios for strong, radiative and hadronic transitions for the 1F and 2F states of $B_c$ mesons (format as in the Table~\ref{results-1}).}
\begin{tabular}{c c c c c c}
\hline\hline
Meson\hspace{.1 in}&\hspace{.1 in} Decay Mode \hspace{.1 in}&\hspace{.1 in} Photon Energy\hspace{.1 in} & Amplitude($\mathcal{M}$) &\hspace{.1 in} $\Gamma_{thy}$\hspace{.1 in} &\hspace{.1 in} B.R \hspace{.1 in}\\
      &       & MeV       &              & MeV                        &($\%$) \\
\hline
$B_c(1 ^3F_2)$  & $B_c(1^3P_0)+\pi\pi$ &  &  & 0.0004~keV & 0.0007\\
                & $B_c(1^3P_2)+\pi\pi$ &  &  & 0.00004~keV & 0.0001\\
                & $B_c(1\;P_1)+\pi\pi$ &  &  & 0.0001~keV & 0.0002\\
                & $B_c(1\;P'_1)+\pi\pi$ &  &  & 0.0002~keV & 0.0004\\
                & $B_c(1^3D_1)\gamma$ & 154.46 & $\langle 1^3D_1|r|1^3F_2\rangle=4.4823$ & 48.41~keV & 85.02\\
                & $B_c(1D_2)\gamma$   & 150.54 & $\langle 1^3D_2|r|1^3F_2\rangle=4.4823$ & 3.32~keV & 5.83 \\
                & $B_c(1D'_2)\gamma$  & 150.54 & $\langle 1^3D_2|r|1^3F_2\rangle=4.4823$ & 4.98~keV & 8.75\\
                & $B_c(1^3D_3)\gamma$ & 148.59 & $\langle 1^3D_3|r|1^3F_2\rangle=4.4823$ & 0.23~keV & 0.4\\
                &total                & &               & 56.94~keV & 100  \\
\hline
$B_c(1\;F_3)$   & $B_c(1^3P_2)+\pi\pi$ &  &  & 0.0001~keV & 0.0002\\
                & $B_c(1\;P_1)+\pi\pi$ &  &  & 0.001~keV & 0.002\\
                & $B_c(1^3D_3)\gamma$ & 145.46 & $\langle 1^3D_3|r|1^3F_3\rangle=4.4823$ & 2.29~keV & 4.59\\
                & $B_c(1\;D_2)\gamma$ & 147.41 & $\langle 1^3D_2|r|1^3F_3\rangle=4.4823$, & 47.64~keV & 95.41   \\
                &                     &        & $\langle 1^1D_2|r|1^1F_3\rangle=4.4769$ \\
                & $B_c(1\;D'_2)\gamma$ &147.41 & $\langle 1^3D_2|r|1^3F_3\rangle=4.4823$, & 0.0001~keV & 0.0002   \\
                &                     &        & $\langle 1^1D_2|r|1^1F_3\rangle=4.4769$ \\
                &total                & &                                   & 49.93~keV & 100  \\
\hline
$B_c(1\;F'_3)$  & $B_c(1^3P_2)+\pi\pi$ &  &  & 0.0001~keV & 0.0002 \\
                & $B_c(1\;P_1)+\pi\pi$ &  &  & $2\times 10^{-5}$~keV & $\sim 0$\\
                & $B_c(1\;P'_1)+\pi\pi$ &  &  & 0.001~keV & 0.002\\
                & $B_c(1^3D_3)\gamma$ & 152.31 & $\langle 1^3D_3|r|1^3F_3\rangle=4.4823$ & 3.51~keV & 6.13\\
                & $B_c(1\;D_2)\gamma$ & 154.26 & $\langle 1^3D_2|r|1^3F_3\rangle=4.4823$, & 0.18~keV & 0.31   \\
                &                     &        & $\langle 1^1D_2|r|1^1F_3\rangle=4.4769$ \\
                & $B_c(1\;D'_2)\gamma$ & 154.26 & $\langle 1^3D_2|r|1^3F_3\rangle=4.4823$, & 53.53~keV &93.55    \\
                &                     &        & $\langle 1^1D_2|r|1^1F_3\rangle=4.4769$ \\
                &total                & &                                   & 57.22~keV & 100  \\
\hline
$B_c(1 ^3F_4)$  & $B_c(1^3P_2)+\pi\pi$ &  &  & 0.0005~keV & 0.001\\
                & $B_c(1^3D_3)\gamma$ & 148.10 & $\langle 1^3D_3|r|1^3F_4\rangle=4.4823$ & 50.8~keV &$\sim 100$ \\
                &total                & &                                   & 50.8~keV & 100  \\
\hline
$B_c(2 ^3F_2)$  &$D B$    && $^1D_2=+0.0277$        & 2.89 & 76.66 \\
                &$D B^*$  && $^3D_2=+0.0206$        & 0.88 & 23.34\\
                &total    &&                        & 3.77     & 100\\
\hline
$B_c(2\;F_3)$   &$D B^*$  && $^3D_3=-0.0205c_F+0.0237s_F$ & 0.01 & 100\\
                &         && $^3G_3=-0.0017c_F-0.0012s_F$ &      &\\
\hline
$B_c(2\;F'_3)$  &$D B^*$  && $^3D_3=+0.0262c_F+0.0227s_F$ & 2.61 & 100\\
                &         && $^3G_3=-0.0019c_F+0.0022s_F$ &      &\\
\hline
$B_c(2 ^3F_4)$  &$D B$    && $^1G_4=+0.0099$        & 0.37 & 97.37\\
                &$D B^*$  && $^3G_4=-0.002$         & 0.01 & 2.63\\
                &total    &&                        & 0.38     & 100\\
\hline\hline
\end{tabular}
\label{results-8}
\end{table}

\section{EXPERIMENTAL SIGNATURES AND SEARCH STRATEGIES}
\label{sec:experimental signatures}

Our calculated masses of $B_c$ states show that there are three $S$-wave, two $P$-wave, two $D$-wave and one
$F$-wave $B_c$ multiplets lying below the $BD$ threshold $(\approx7144)$ MeV. These are the
narrow states of $B_c$ spectrum because $B_c$ cannot annihilate into
gluons due to its non-zero flavor. All these excited states below the $BD$
threshold will cascade decay into the ground state $B_c$ through
emission of photons and/or pions, which eventually decays through weak interaction.
These photons and pions produced by electromagnetic and/or hadronic transitions carry unique signature of initial $B_c$. Hence the production events of $B_c$ excited states can be reconstructed by detecting and measuring the energies of produced photons and pions.
On the other hand the excited states above the $BD$
threshold will rapidly decay into a pair of $B(B_s)$ and $D(D_s)$
mesons through strong interaction processes. In order to predict observable event rates of $B_c$ excited states (below $BD$ threshold) in pp collision at LHC, we require the knowledge of their production cross sections, branching ratios of their electromagnetic and hadronic transitions, and the branching ratios of weak decay channels of $B_c$ ground states through which its production can be identified experimentally.

Inclusive production cross sections of $B_c$ states in pp collision at LHC energy has been calculated in Refs. \cite{cheung-1993,cheung-1994,cheung-1996p,cheung-1996d} using fragmentation approach and in Refs. \cite{chang-1993,chang-1995,chang-1996,chang-2004,kolod-1995,bere-1997} using pQCD approach. The results of two approaches quantitatively agree for $p_T\geq10$ GeV \cite{kolod-1995}. Fragmentation approach of Refs. \cite{cheung-1993,cheung-1994} shows that production cross sections of $B_c(1^1S_0)$ and $B_c(1^3S_1)$ states are 0.72 and 1.21 nb respectively for transverse momentum $p_T(B_c)>20$ GeV and rapidity $|y(B_c)|<2.5$ at LHC. These values include the contribution of both $\bar{b}$-quark and gluon fragmentation functions of the $B_c$ states. When these values are extrapolated to kinematic cut $p_T(B_c)>10$ GeV using the values reported in Table III of Ref. \cite{cheung-1996p}, we obtain the production cross sections 5.5 and 9.3 nb for the $1^1S_0$ and $1^3S_1$ states respectively. The production cross sections of the $2^1S_0$ and $2^3S_1$ states are obtained by multiplying the corresponding values of $1S$ states with the factor $|R_{2S}(0)/R_{1S}(0)|^2\simeq0.6$ \cite{cheung-1993}. In Ref. \cite{cheung-1996p} production cross sections of $1P$ and $2P$ states are calculated using the fragmentation approach. However, for LHC they report only total cross section that include the contribution from 1S, 2S, 1P and 2P states. The reported value 33.8 nb for kinematic cuts $p_T(B_c)>10$ GeV and $|y(B_c)|<2.5$ implies that total contribution of 1P and 2P states is 10.2 nb. We divide this value over eight 1P and 2P states using the distribution reported for Tevatron in Fig. 3 and 4 of Ref. \cite{cheung-1996p}. It is pointed in Ref. \cite{cheung-1994} that the distribution is not much changed at the LHC energy. In Ref. \cite{cheung-1996d}, it is shown that total fragmentation probability for a $\bar{b}$-quark to produce the D-wave $B_c$ mesons is about $2\times10^{-5}$, equivalent to 2\% of the total inclusive cross section of all of $B_c$ states lying below $BD$ threshold.  These estimates of the cross sections are used to predict the number of events of various decay chains of excited $B_c$ states.

As the excited states below the $BD$ threshold eventually decay
into the $B_c$ ground state, therefore it is important to observe
this state in order to reconstruct the events of originally produced states. Prominent weak decay modes of $B_c$ ground states are summarized in Table 10 of Ref. \cite{godfrey-2004-Bc}. We assume that
ground state $B_c$ is observed through two golden channels:
i) $B_c^{\pm} \rightarrow J/\psi+\pi^{\pm} \rightarrow l^-l^+
\pi^{\pm}$ having combined BR of $0.013\%$ and detection
efficiency of $\sim 2\%$ and ii) $B_c^{\pm} \rightarrow J/\psi
l^{\pm}\nu_l \rightarrow l'^-l'^+ l^{\pm}\nu_l$ having combined BR of
$0.21\%$ and detection efficiency of $\sim 4\%$~(See
Table~\ref{branching ratios} for the branching ratios).
\begin{table}[h!]
\centering
\caption{Branching ratios for the two golden channels of $B_c$ along the branching ratio of $J/\psi\rightarrow l^+l^-$.}
\begin{tabular}{c c c}
\hline\hline
Decay Process\hspace{.1 in}&\hspace{.1 in} Branching Ratio~($\%$)\\
\hline
$B_c^{\pm} \rightarrow J/\psi \pi^{\pm}$ &  $0.111^{+0.009}_{-0.010}$~\cite{chang-2015}  \\
$B_c^{\pm} \rightarrow J/\psi l^{\pm}\nu_l$ &  $1.73\pm 0.05$~\cite{chang-2015}  \\
\hline
$J/\psi\rightarrow l^+l^-$ & $11.9\pm0.06$~\cite{PDG-2016}\\
\hline\hline
\end{tabular}
\label{branching ratios}
\end{table}
We calculate the number of events of various decay chains of excited $B_c$ states below $BD$ threshold at LHC for integrated luminosity $L=100~\text{fb}^{-1}$. The values reported in Tables \ref{decay-chains-1} - \ref{decay-chains-7} include the events observed through both of the golden channels. The decay chains having yield less than 100 are not included in these tables. We include a factor of 2 to incorporate both charge conjugate states of $B_c$.
It is noted that our mass calculations show that $3S$, $2P$, $2D$, and $1F$ states are below but close to $BD$ threshold (energy gap $ < 0.15$ GeV). These results significantly differ from the mass predictions given in Refs. \cite{ferretti-2018,ebert-2003,godfrey-2004-Bc}. Ref. \cite{ferretti-2018} shows that $3S$, $2^3P_2$, $2D$, and $1F$ states are above $BD$ threshold, whereas Ref. \cite{godfrey-2004-Bc} shows that three multiplets of $2P$ states are above $BD$ threshold along with $3S$, $2D$, and $1F$ states. Ref. \cite{ebert-2003} also shows that $3S$ and three multiplets of $2P$ states are above $BD$ threshold. The result is that these states, that are expected to be observed through radiative and hadronic transition according to our mass predictions, are unlikely to be observed in these references. This also gives significantly different branching ratios and predicted strong decay widths. Thus the experimenters should treat our predictions of branching ratios and strong decay widths of the states close to $BD$ threshold cautiously.  Besides this caveat, there are no results available for the production cross sections of $3S$, $2D$, and $1F$  states in pp collision, therefore,  we abstain to make any predictions of event rates for these states.

These results show that in LHC it is possible to produce sufficient number of events corresponding to different decay chains of the excited $B_c$ states
below $BD$ threshold. The task of event reconstruction become much easier when an excited $B_c$ state directly decays to the ground state through E1/M1 or hadronic transitions. Tables \ref{results-1} and \ref{results-2} show that $1 ^3S_1$, $2 ^3S_1$, $1P_1$, $1P'_1$, $2P_1$, $2P'_1$ states can directly decay to $B_c$ ground states through E1/M1 transitions.  All these direct transitions also appear in the tables \ref{decay-chains-1}, \ref{decay-chains-2}, \ref{decay-chains-4} and \ref{decay-chains-5} of decay chains as their yield is much higher than 100. The case of $2 ^3S_1\rightarrow1 ^1S_0+\gamma$ is particulary interesting. Only 2650 events are expected in this case owing to small value of its BR $(\approx2.75\%)$ despite having relatively large production cross section of $2 ^3S_1$ state. Therefore, the best way to search this state is via $2 ^3S_1\xlongrightarrow{\text{$\gamma$}}1 ^3P_2
\xlongrightarrow{\text{$\gamma$}}1 ^3S_1\xlongrightarrow{\text{$\gamma$}}1 ^1S_0$ or $2 ^3S_1\xlongrightarrow{\text{$\pi\pi$}}1 ^3S_1\xlongrightarrow{\text{$\gamma$}}1 ^1S_0$ for which  expected number of events are $2.78 \times10^4$ and $2.58 \times10^4$ respectively as shown in the Table \ref{decay-chains-1}. Tables \ref{results-1} and \ref{results-4} show that $2 ^1S_0$, $1D_2$, and $1D'_2$ states can directly decay to $B_c$ ground state through hadronic transitions. However, due to small BRs of hardonic transitions of $1D_2$ and $1D'_2$ (less than $1\%$), the resultant number of events are less than 100 and are not included in the Table \ref{decay-chains-7}. The best way to detect $D$ states is via double or triple photon emission as given in the Table \ref{decay-chains-7}. Our results given in the tables \ref{decay-chains-1}-\ref{decay-chains-7} of decay chains can help experimentalists in adopting the best strategies to discover and study properties of the excited $B_c$ states below $BD$ threshold.

\begin{table}[h!]
\centering
\caption{Decay chains of $1S$ and $2S$ states and expected number of events at LHC.}
\begin{tabular}{c c c}
\hline\hline
Initial state & Decay Chain\hspace{.1 in}&\hspace{.1 in} Number of events \hspace{.1 in}\\
\hline
$1^3S_1$ & $\xlongrightarrow[\text{100$\%$}]{\text{$\gamma$}} B_c   $         & $1.6\times 10^5$  \\
$2^1S_0$ & $\xlongrightarrow[\text{51.42$\%$}]{\text{$\pi\pi$}} B_c$         & $2.96\times 10^4$ \\
         & $\xlongrightarrow[\text{34.23$\%$}]{\text{$\gamma$}}1\;P_1 \xlongrightarrow[\text{68.78$\%$}]{\text{$\gamma$}} B_c$ & $1.35\times 10^4$   \\
         & $\xlongrightarrow[\text{34.23$\%$}]{\text{$\gamma$}}1\;P_1 \xlongrightarrow[\text{31.22$\%$}]{\text{$\gamma$}}1^3S_1\xlongrightarrow[\text{100$\%$}]{\text{$\gamma$}} B_c$ & $6.15\times 10^3$   \\
         & $\xlongrightarrow[\text{14.01$\%$}]{\text{$\gamma$}}1\;P'_1 \xlongrightarrow[\text{35.56$\%$}]{\text{$\gamma$}} B_c$    & $2.86\times 10^3$   \\
         & $\xlongrightarrow[\text{14.01$\%$}]{\text{$\gamma$}}1\;P'_1 \xlongrightarrow[\text{64.44$\%$}]{\text{$\gamma$}}1^3S_1\xlongrightarrow[\text{100$\%$}]{\text{$\gamma$}} B_c$         & $5.19\times 10^3$   \\
$2^3S_1$ & $\xlongrightarrow[\text{26.76$\%$}]{\text{$\pi\pi$}}1^3S_1 \xlongrightarrow[\text{100$\%$}]{\text{$\gamma$}} B_c$         & $2.58\times 10^4$  \\
         & $\xlongrightarrow[\text{28.91$\%$}]{\text{$\gamma$}}1^3P_2 \xlongrightarrow[\text{100$\%$}]{\text{$\gamma$}}1^3S_1\xlongrightarrow[\text{100$\%$}]{\text{$\gamma$}} B_c$         & $2.78\times 10^4$   \\
         & $\xlongrightarrow[\text{2.75$\%$}]{\text{$\gamma$}} B_c$         & $2.65\times 10^3$  \\
         & $\xlongrightarrow[\text{9.59$\%$}]{\text{$\gamma$}}1\;P_1 \xlongrightarrow[\text{68.78$\%$}]{\text{$\gamma$}} B_c$         & $6.35\times 10^3$  \\
         & $\xlongrightarrow[\text{9.59$\%$}]{\text{$\gamma$}}1\;P_1 \xlongrightarrow[\text{31.22$\%$}]{\text{$\gamma$}}1^3S_1\xlongrightarrow[\text{100$\%$}]{\text{$\gamma$}} B_c$         & $2.88\times 10^3$   \\
         & $\xlongrightarrow[\text{15.76$\%$}]{\text{$\gamma$}}1\;P'_1 \xlongrightarrow[\text{35.56$\%$}]{\text{$\gamma$}} B_c$         & $5.4\times 10^3$  \\
         & $\xlongrightarrow[\text{15.76$\%$}]{\text{$\gamma$}}1\;P'_1 \xlongrightarrow[\text{64.44$\%$}]{\text{$\gamma$}}1^3S_1\xlongrightarrow[\text{100$\%$}]{\text{$\gamma$}} B_c$         & $9.78\times 10^3$   \\
         & $\xlongrightarrow[\text{16.23$\%$}]{\text{$\gamma$}}1^3P_0 \xlongrightarrow[\text{100$\%$}]{\text{$\gamma$}}1^3S_1\xlongrightarrow[\text{100$\%$}]{\text{$\gamma$}} B_c$         & $1.56\times 10^4$   \\
\hline\hline
\end{tabular}
\label{decay-chains-1}
\end{table}

\begin{table}[h!]
\centering
\caption{Decay chains of $1P$ states and expected number of events at LHC.}
\begin{tabular}{c c c}
\hline\hline
Initial state & Decay Chain\hspace{.1 in}&\hspace{.1 in} Number of events \hspace{.1 in}\\
\hline
$1^3P_2$   & $\xlongrightarrow[\text{100$\%$}]{\text{$\gamma$}}1^3S_1\xlongrightarrow[\text{100$\%$}]{\text{$\gamma$}} B_c$    & $2.6\times 10^4$ \\
$1P^{'}_1$ & $\xlongrightarrow[\text{35.56$\%$}]{\text{$\gamma$}} B_c$         & $6.53\times 10^3$ \\
           & $\xlongrightarrow[\text{64.44$\%$}]{\text{$\gamma$}}1^3S_1 \xlongrightarrow[\text{100$\%$}]{\text{$\gamma$}} B_c$    & $1.18\times 10^4$   \\
$1P_1$     & $\xlongrightarrow[\text{31.22$\%$}]{\text{$\gamma$}}1^3S_1\xlongrightarrow[\text{100$\%$}]{\text{$\gamma$}} B_c$      & $5.73\times 10^3$   \\
           & $\xlongrightarrow[\text{68.78$\%$}]{\text{$\gamma$}} B_c$         & $1.26\times 10^4$  \\
$1^3P_0$   & $\xlongrightarrow[\text{100$\%$}]{\text{$\gamma$}}1^3S_1\xlongrightarrow[\text{100$\%$}]{\text{$\gamma$}} B_c$       & $1.23\times 10^4$   \\
\hline\hline
\end{tabular}
\label{decay-chains-2}
\end{table}

\begin{table}[h!]
\centering
\caption{Decay chains of $2^3P_2$ states and expected number of events at LHC.}
\begin{tabular}{c c c}
\hline\hline
Initial state & Decay Chain\hspace{.1 in}&\hspace{.1 in} Number of events \hspace{.1 in}\\
\hline
$2^3P_2$   & $\xlongrightarrow[\text{27.03$\%$}]{\text{$\gamma$}}1^3S_1\xlongrightarrow[\text{100$\%$}]{\text{$\gamma$}}B_c$    & $9.08\times 10^3$ \\
           & $\xlongrightarrow[\text{59.31$\%$}]{\text{$\gamma$}}2^3S_1\xlongrightarrow[\text{26.76$\%$}]{\text{$\pi\pi$}}
           1^3S_1\xlongrightarrow[\text{100$\%$}]{\text{$\gamma$}}B_c$    & $5.33\times 10^3$ \\
           & $\xlongrightarrow[\text{59.31$\%$}]{\text{$\gamma$}}2^3S_1\xlongrightarrow[\text{28.91$\%$}]{\text{$\gamma$}}
           1^3P_2\xlongrightarrow[\text{100$\%$}]{\text{$\gamma$}}1^3S_1\xlongrightarrow[\text{100$\%$}]{\text{$\gamma$}}B_c$    & $5.76\times 10^3$ \\
           & $\xlongrightarrow[\text{59.31$\%$}]{\text{$\gamma$}}2^3S_1\xlongrightarrow[\text{9.59$\%$}]{\text{$\gamma$}}
           1\;P_1\xlongrightarrow[\text{68.78$\%$}]{\text{$\gamma$}}B_c$    & $1.31\times 10^3$ \\
           & $\xlongrightarrow[\text{59.31$\%$}]{\text{$\gamma$}}2^3S_1\xlongrightarrow[\text{15.76$\%$}]{\text{$\gamma$}}
           1\;P'_1\xlongrightarrow[\text{35.56$\%$}]{\text{$\gamma$}}B_c$    & $1.12\times 10^3$ \\
           & $\xlongrightarrow[\text{59.31$\%$}]{\text{$\gamma$}}2^3S_1\xlongrightarrow[\text{15.76$\%$}]{\text{$\gamma$}}
           1\;P'_1\xlongrightarrow[\text{64.44$\%$}]{\text{$\gamma$}}1^3S_1 \xlongrightarrow[\text{100$\%$}]{\text{$\gamma$}}B_c$    & $2.02\times 10^3$ \\
           & $\xlongrightarrow[\text{59.31$\%$}]{\text{$\gamma$}}2^3S_1\xlongrightarrow[\text{16.23$\%$}]{\text{$\gamma$}}
           1^3P_0\xlongrightarrow[\text{100$\%$}]{\text{$\gamma$}}1^3S_1 \xlongrightarrow[\text{100$\%$}]{\text{$\gamma$}}B_c$    & $3.23\times 10^3$ \\
           & $\xlongrightarrow[\text{11.34$\%$}]{\text{$\gamma$}}1^3D_3\xlongrightarrow[\text{99.91$\%$}]{\text{$\gamma$}}
           1^3P_2\xlongrightarrow[\text{100$\%$}]{\text{$\gamma$}}1^3S_1 \xlongrightarrow[\text{100$\%$}]{\text{$\gamma$}}B_c$    & $3.81\times 10^3$ \\
           & $\xlongrightarrow[\text{0.86$\%$}]{\text{$\gamma$}}1\;D_2\xlongrightarrow[\text{91.8$\%$}]{\text{$\gamma$}}
           1\;P_1\xlongrightarrow[\text{68.78$\%$}]{\text{$\gamma$}}B_c$    & $1.82\times 10^2$ \\
           & $\xlongrightarrow[\text{1.29$\%$}]{\text{$\gamma$}}1\;D'_2\xlongrightarrow[\text{84.86$\%$}]{\text{$\gamma$}}
           1\;P'_1\xlongrightarrow[\text{35.56$\%$}]{\text{$\gamma$}}B_c$    & $1.31\times 10^2$ \\
           & $\xlongrightarrow[\text{1.29$\%$}]{\text{$\gamma$}}1\;D'_2\xlongrightarrow[\text{84.86$\%$}]{\text{$\gamma$}}
           1\;P'_1\xlongrightarrow[\text{64.44$\%$}]{\text{$\gamma$}}1^3S_1\xlongrightarrow[\text{100$\%$}]{\text{$\gamma$}}B_c$    & $2.37\times 10^2$ \\
\hline\hline
\end{tabular}
\label{decay-chains-3}
\end{table}

\begin{table}[h!]
\centering
\caption{Decay chains of $2P'_1$ states and expected number of events at LHC.}
\begin{tabular}{c c c}
\hline\hline
Initial state & Decay Chain\hspace{.1 in}&\hspace{.1 in} Number of events \hspace{.1 in}\\
\hline
$2\;P'_1$  & $\xlongrightarrow[\text{10.83$\%$}]{\text{$\gamma$}} B_c$      & $3.64\times 10^3$   \\
           & $\xlongrightarrow[\text{20.08$\%$}]{\text{$\gamma$}}2^1S_0 \xlongrightarrow[\text{51.42$\%$}]{\text{$\gamma$}} B_c$     & $3.47\times 10^3$  \\
           & $\xlongrightarrow[\text{20.08$\%$}]{\text{$\gamma$}}2^1S_0 \xlongrightarrow[\text{34.23$\%$}]{\text{$\gamma$}}1\;P_1\xlongrightarrow[\text{68.78$\%$}]{\text{$\gamma$}} B_c$       & $1.59\times 10^3$   \\
           & $\xlongrightarrow[\text{20.08$\%$}]{\text{$\gamma$}}2^1S_0 \xlongrightarrow[\text{34.23$\%$}]{\text{$\gamma$}}1\;P_1 \xlongrightarrow[\text{31.22$\%$}]{\text{$\gamma$}} 1^3S_1 \xlongrightarrow[\text{100$\%$}]{\text{$\gamma$}} B_c$       & $7.2\times 10^2$   \\
           & $\xlongrightarrow[\text{20.08$\%$}]{\text{$\gamma$}}2^1S_0 \xlongrightarrow[\text{14.01$\%$}]{\text{$\gamma$}}1\;P'_1 \xlongrightarrow[\text{64.44$\%$}]{\text{$\gamma$}} 1^3S_1 \xlongrightarrow[\text{100$\%$}]{\text{$\gamma$}} B_c$       & $6.09\times 10^2$   \\
           & $\xlongrightarrow[\text{37.87$\%$}]{\text{$\gamma$}}2^3S_1 \xlongrightarrow[\text{26.76$\%$}]{\text{$\pi\pi$}}1^3S_1 \xlongrightarrow[\text{100$\%$}]{\text{$\gamma$}} B_c$       & $3.41\times 10^3$   \\
           & $\xlongrightarrow[\text{37.87$\%$}]{\text{$\gamma$}}2^3S_1 \xlongrightarrow[\text{28.91$\%$}]{\text{$\gamma$}}1^3P_2 \xlongrightarrow[\text{100$\%$}]{\text{$\gamma$}}1^3S_1 \xlongrightarrow[\text{100$\%$}]{\text{$\gamma$}} B_c$       & $3.68\times 10^3$   \\
           & $\xlongrightarrow[\text{37.87$\%$}]{\text{$\gamma$}}2^3S_1 \xlongrightarrow[\text{9.59$\%$}]{\text{$\gamma$}}1\;P_1 \xlongrightarrow[\text{68.78$\%$}]{\text{$\gamma$}} B_c$       & $8.39\times 10^2$   \\
           & $\xlongrightarrow[\text{37.87$\%$}]{\text{$\gamma$}}2^3S_1 \xlongrightarrow[\text{15.76$\%$}]{\text{$\gamma$}}1\;P'_1 \xlongrightarrow[\text{64.44$\%$}]{\text{$\gamma$}}1^3S_1 \xlongrightarrow[\text{100$\%$}]{\text{$\gamma$}}B_c$       & $1.29\times 10^3$   \\
           & $\xlongrightarrow[\text{37.87$\%$}]{\text{$\gamma$}}2^3S_1 \xlongrightarrow[\text{16.23$\%$}]{\text{$\gamma$}}1^3P_0 \xlongrightarrow[\text{100$\%$}]{\text{$\gamma$}}1^3S_1 \xlongrightarrow[\text{100$\%$}]{\text{$\gamma$}}B_c$       & $2.07\times 10^3$   \\
           & $\xlongrightarrow[\text{2.32$\%$}]{\text{$\gamma$}}1^3D_1 \xlongrightarrow[\text{63.97$\%$}]{\text{$\gamma$}}1^3P_0 \xlongrightarrow[\text{100$\%$}]{\text{$\gamma$}}1^3S_1 \xlongrightarrow[\text{100$\%$}]{\text{$\gamma$}}B_c$       & $4.99\times 10^2$   \\
           & $\xlongrightarrow[\text{2.32$\%$}]{\text{$\gamma$}}1^3D_1 \xlongrightarrow[\text{21.98$\%$}]{\text{$\gamma$}}1\;P'_1 \xlongrightarrow[\text{64.44$\%$}]{\text{$\gamma$}}1^3S_1 \xlongrightarrow[\text{100$\%$}]{\text{$\gamma$}}B_c$       & $1.1\times 10^2$   \\
           & $\xlongrightarrow[\text{10.33$\%$}]{\text{$\gamma$}}1\;D'_2 \xlongrightarrow[\text{13.32$\%$}]{\text{$\gamma$}}1^3P_2 \xlongrightarrow[\text{100$\%$}]{\text{$\gamma$}}1^3S_1 \xlongrightarrow[\text{100$\%$}]{\text{$\gamma$}}B_c$       & $4.62\times 10^2$   \\
           & $\xlongrightarrow[\text{10.33$\%$}]{\text{$\gamma$}}1\;D'_2 \xlongrightarrow[\text{84.86$\%$}]{\text{$\gamma$}}1\;P'_1 \xlongrightarrow[\text{35.56$\%$}]{\text{$\gamma$}}B_c$       & $1.05\times 10^3$   \\
           & $\xlongrightarrow[\text{10.33$\%$}]{\text{$\gamma$}}1\;D'_2 \xlongrightarrow[\text{84.86$\%$}]{\text{$\gamma$}}1\;P'_1 \xlongrightarrow[\text{64.44$\%$}]{\text{$\gamma$}}1^3S_1 \xlongrightarrow[\text{100$\%$}]{\text{$\gamma$}}B_c$       & $1.9\times 10^3$   \\
\hline\hline
\end{tabular}
\label{decay-chains-4}
\end{table}

\begin{table}[h!]
\centering
\caption{Decay chains of $2\;P_1$ states and expected number of events at LHC.}
\begin{tabular}{c c c}
\hline\hline
Initial state & Decay Chain\hspace{.1 in}&\hspace{.1 in} Number of events \hspace{.1 in}\\
\hline
$2\;P_1$   & $\xlongrightarrow[\text{23.16$\%$}]{\text{$\gamma$}} B_c$      & $4.25\times 10^3$   \\
           & $\xlongrightarrow[\text{9.97$\%$}]{\text{$\gamma$}}1^3S_1 \xlongrightarrow[\text{100$\%$}]{\text{$\gamma$}} B_c$      & $1.83\times 10^3$   \\
           & $\xlongrightarrow[\text{39.02$\%$}]{\text{$\gamma$}}2^1S_0 \xlongrightarrow[\text{51.42$\%$}]{\text{$\pi\pi$}} B_c$     & $3.68\times 10^3$  \\
           & $\xlongrightarrow[\text{39.02$\%$}]{\text{$\gamma$}}2^1S_0 \xlongrightarrow[\text{34.23$\%$}]{\text{$\gamma$}}1\;P_1\xlongrightarrow[\text{68.78$\%$}]{\text{$\gamma$}} B_c$       & $1.69\times 10^3$   \\
           & $\xlongrightarrow[\text{39.02$\%$}]{\text{$\gamma$}}2^1S_0 \xlongrightarrow[\text{34.23$\%$}]{\text{$\gamma$}}1\;P_1 \xlongrightarrow[\text{31.22$\%$}]{\text{$\gamma$}} 1^3S_1 \xlongrightarrow[\text{100$\%$}]{\text{$\gamma$}} B_c$       & $7.66\times 10^2$   \\
           & $\xlongrightarrow[\text{39.02$\%$}]{\text{$\gamma$}}2^1S_0 \xlongrightarrow[\text{14.01$\%$}]{\text{$\gamma$}}1\;P'_1 \xlongrightarrow[\text{64.44$\%$}]{\text{$\gamma$}} 1^3S_1 \xlongrightarrow[\text{100$\%$}]{\text{$\gamma$}} B_c$       & $6.47\times 10^2$   \\
           & $\xlongrightarrow[\text{18.41$\%$}]{\text{$\gamma$}}2^3S_1 \xlongrightarrow[\text{26.76$\%$}]{\text{$\pi\pi$}}1^3S_1 \xlongrightarrow[\text{100$\%$}]{\text{$\gamma$}} B_c$       & $9.04\times 10^2$   \\
           & $\xlongrightarrow[\text{18.41$\%$}]{\text{$\gamma$}}2^3S_1 \xlongrightarrow[\text{28.91$\%$}]{\text{$\gamma$}}1^3P_2 \xlongrightarrow[\text{100$\%$}]{\text{$\gamma$}}1^3S_1 \xlongrightarrow[\text{100$\%$}]{\text{$\gamma$}} B_c$       & $9.77\times 10^2$   \\
           & $\xlongrightarrow[\text{9.27$\%$}]{\text{$\gamma$}}1\;D_2 \xlongrightarrow[\text{8.12$\%$}]{\text{$\gamma$}}1^3P_2 \xlongrightarrow[\text{100$\%$}]{\text{$\gamma$}}1^3S_1 \xlongrightarrow[\text{100$\%$}]{\text{$\gamma$}}B_c$       & $1.38\times 10^2$   \\
           & $\xlongrightarrow[\text{9.27$\%$}]{\text{$\gamma$}}1\;D_2 \xlongrightarrow[\text{91.8$\%$}]{\text{$\gamma$}}1\;P_1 \xlongrightarrow[\text{68.78$\%$}]{\text{$\gamma$}}B_c$       & $1.07\times 10^3$   \\
           & $\xlongrightarrow[\text{9.27$\%$}]{\text{$\gamma$}}1\;D_2 \xlongrightarrow[\text{91.8$\%$}]{\text{$\gamma$}}1\;P_1 \xlongrightarrow[\text{31.22$\%$}]{\text{$\gamma$}}1^3S_1 \xlongrightarrow[\text{100$\%$}]{\text{$\gamma$}}B_c$       & $4.88\times 10^2$   \\
\hline\hline
\end{tabular}
\label{decay-chains-5}
\end{table}

\begin{table}[h!]
\centering
\caption{Decay chains of $2^3P_0$ states  and expected number of events at LHC.}
\begin{tabular}{c c c}
\hline\hline
Initial state & Decay Chain\hspace{.1 in}&\hspace{.1 in} Number of events \hspace{.1 in}\\
\hline
$2^3P_0$   & $\xlongrightarrow[\text{35.87$\%$}]{\text{$\gamma$}}1^3S_1\xlongrightarrow[\text{100$\%$}]{\text{$\gamma$}}B_c$    & $5.47\times 10^3$ \\
           & $\xlongrightarrow[\text{55.38$\%$}]{\text{$\gamma$}}2^3S_1\xlongrightarrow[\text{26.76$\%$}]{\text{$\pi\pi$}}
           1^3S_1\xlongrightarrow[\text{100$\%$}]{\text{$\gamma$}}B_c$    & $2.26\times 10^3$ \\
           & $\xlongrightarrow[\text{55.38$\%$}]{\text{$\gamma$}}2^3S_1\xlongrightarrow[\text{28.91$\%$}]{\text{$\gamma$}}
           1^3P_2\xlongrightarrow[\text{100$\%$}]{\text{$\gamma$}}1^3S_1\xlongrightarrow[\text{100$\%$}]{\text{$\gamma$}}B_c$    & $2.44\times 10^3$ \\
           & $\xlongrightarrow[\text{55.38$\%$}]{\text{$\gamma$}}2^3S_1\xlongrightarrow[\text{15.76$\%$}]{\text{$\gamma$}} 1\;P'_1 \xlongrightarrow[\text{64.44$\%$}]{\text{$\gamma$}}1^3S_1\xlongrightarrow[\text{100$\%$}]{\text{$\gamma$}}B_c$    & $8.57\times 10^2$ \\
           & $\xlongrightarrow[\text{55.38$\%$}]{\text{$\gamma$}}2^3S_1\xlongrightarrow[\text{16.23$\%$}]{\text{$\gamma$}} 1^3P_0 \xlongrightarrow[\text{100$\%$}]{\text{$\gamma$}}1^3S_1\xlongrightarrow[\text{100$\%$}]{\text{$\gamma$}}B_c$    & $1.37\times 10^3$ \\
           & $\xlongrightarrow[\text{8.74$\%$}]{\text{$\gamma$}}1^3D_1\xlongrightarrow[\text{63.97$\%$}]{\text{$\gamma$}}
           1^3P_0\xlongrightarrow[\text{100$\%$}]{\text{$\gamma$}}1^3S_1\xlongrightarrow[\text{100$\%$}]{\text{$\gamma$}}B_c$    & $8.52\times 10^2$ \\
           & $\xlongrightarrow[\text{8.74$\%$}]{\text{$\gamma$}}1^3D_1\xlongrightarrow[\text{12.1$\%$}]{\text{$\gamma$}}
           1\;P_1\xlongrightarrow[\text{68.78$\%$}]{\text{$\gamma$}}B_c$    & $1.61\times 10^2$ \\
           & $\xlongrightarrow[\text{8.74$\%$}]{\text{$\gamma$}}1^3D_1\xlongrightarrow[\text{21.98$\%$}]{\text{$\gamma$}}
           1\;P'_1\xlongrightarrow[\text{64.44$\%$}]{\text{$\gamma$}}1^3S_1\xlongrightarrow[\text{100$\%$}]{\text{$\gamma$}}B_c$    & $1.89\times 10^2$ \\
\hline\hline
\end{tabular}
\label{decay-chains-6}
\end{table}

\begin{table}[h!]
\centering
\caption{Decay chains of $1D$ states and expected number of events at LHC.}
\begin{tabular}{c c c}
\hline\hline
Initial state & Decay Chain\hspace{.1 in}&\hspace{.1 in} Number of events \hspace{.1 in}\\
\hline
$1^3D_3$   & $\xlongrightarrow[\text{99.91$\%$}]{\text{$\gamma$}}1^3P_2\xlongrightarrow[\text{100$\%$}]{\text{$\gamma$}}1^3S_1\xlongrightarrow[\text{100$\%$}]{\text{$\gamma$}} B_c$    & $2.92\times 10^3$ \\
$1\;D'_2$  & $\xlongrightarrow[\text{13.32$\%$}]{\text{$\gamma$}}1^3P_2 \xlongrightarrow[\text{100$\%$}]{\text{$\gamma$}}1^3S_1
            \xlongrightarrow[\text{100$\%$}]{\text{$\gamma$}} B_c$       & $3.9\times 10^2$ \\
           & $\xlongrightarrow[\text{84.86$\%$}]{\text{$\gamma$}}1\;P'_1 \xlongrightarrow[\text{35.56$\%$}]{\text{$\gamma$}} B_c$    & $8.83\times 10^2$  \\
           & $\xlongrightarrow[\text{84.86$\%$}]{\text{$\gamma$}}1\;P'_1 \xlongrightarrow[\text{64.44$\%$}]{\text{$\gamma$}}1^3S_1\xlongrightarrow[\text{100$\%$}]{\text{$\gamma$}} B_c$    & $1.6\times 10^3$  \\
$1\;D_2$   & $\xlongrightarrow[\text{8.12$\%$}]{\text{$\gamma$}}1^3P_2\xlongrightarrow[\text{100$\%$}]{\text{$\gamma$}} 1^3S_1 \xlongrightarrow[\text{100$\%$}]{\text{$\gamma$}} B_c$      & $2.38\times 10^2$   \\
           & $\xlongrightarrow[\text{91.8$\%$}]{\text{$\gamma$}}1\;P_1 \xlongrightarrow[\text{68.78$\%$}]{\text{$\gamma$}} B_c$  & $1.85\times 10^3$  \\
           & $\xlongrightarrow[\text{91.8$\%$}]{\text{$\gamma$}}1\;P_1 \xlongrightarrow[\text{31.22$\%$}]{\text{$\gamma$}}1^3S_1 \xlongrightarrow[\text{100$\%$}]{\text{$\gamma$}} B_c$  & $8.39\times 10^2$  \\
$1^3D_1$   & $\xlongrightarrow[\text{63.97$\%$}]{\text{$\gamma$}}1^3P_0\xlongrightarrow[\text{100$\%$}]{\text{$\gamma$}} 1^3S_1 \xlongrightarrow[\text{100$\%$}]{\text{$\gamma$}} B_c$       & $1.87\times 10^3$   \\
           & $\xlongrightarrow[\text{12.1$\%$}]{\text{$\gamma$}}1\;P_1\xlongrightarrow[\text{68.78$\%$}]{\text{$\gamma$}} B_c$  & $2.45\times 10^2$  \\
           & $\xlongrightarrow[\text{12.1$\%$}]{\text{$\gamma$}}1\;P_1\xlongrightarrow[\text{31.22$\%$}]{\text{$\gamma$}} 1^3S_1 \xlongrightarrow[\text{100$\%$}]{\text{$\gamma$}} B_c$  & $1.11\times 10^2$  \\
           & $\xlongrightarrow[\text{21.98$\%$}]{\text{$\gamma$}}1\;P'_1\xlongrightarrow[\text{35.56$\%$}]{\text{$\gamma$}} B_c$  & $2.29\times 10^2$  \\
           & $\xlongrightarrow[\text{21.98$\%$}]{\text{$\gamma$}}1\;P'_1\xlongrightarrow[\text{64.44$\%$}]{\text{$\gamma$}} 1^3S_1 \xlongrightarrow[\text{100$\%$}]{\text{$\gamma$}} B_c$  & $4.15\times 10^2$  \\
\hline\hline
\end{tabular}
\label{decay-chains-7}
\end{table}

\section{Concluding Remarks}
\label{conclusions}

In this paper we studied the properties of charmed-bottom mesons including masses, radiative transitions, hadronic transitions and the OZI allowed strong decays. We have computed the spectrum of $B_c$ mesons upto $2F$ states with a non-relativistic quark model that incorporates scalar confinement and one gluon exchange spin-dependent interactions. These eigenfunctions were then used to obtain E1 and M1 radiative transitions. Strong decay amplitudes of excited $B_c$ states above the $BD$ threshold have been obtained using the modified $^3P_0$ pair creation model and fitted SHO wave functions. The hadronic transition rates for $B_c$ mesons have been predicted using the Kuang-Yan approach. The total decay widths of excited $B_c$ states have been predicted by summing the radiative, hadronic and strong widths. The branching ratios of different final states are estimated by using the total widths. These branching ratios are then combined with production rates at the LHC to estimate the number of events of various decay chains of excited $B_c$ states.
We expect that the predictions presented in this work will be help experimentalists find the excited $B_c$ states at LHC and measure their properties.

\section{Acknowledgement}
FA acknowledges the financial support of HEC of Pakistan through Project: 20-4500/NRPU/R\&D/HEC/14/727.


\end{document}